\documentclass[10pt,journal,compsoc]{IEEEtran}

\usepackage[ruled,vlined,linesnumbered]{algorithm2e}
\usepackage[final,inline,nomargin,index]{fixme}
\fxsetup{theme=color,mode=multiuser}
\usepackage[table]{xcolor}
\usepackage{algpseudocode}
\usepackage{amssymb}
\usepackage{amsthm}
\usepackage{amsmath}
\usepackage{balance}
\usepackage{booktabs}
\usepackage{graphicx}
\usepackage[]{hyperref}
\hypersetup{
  colorlinks,
  linkcolor={blue!50!black},
  citecolor={blue!50!black},
  urlcolor={blue!50!black},
  filecolor={blue!50!black}
}
\usepackage{inconsolata}
\usepackage{multirow}
\usepackage{pifont}
\usepackage{subcaption}
\usepackage{tabularx}
\usepackage{threeparttable}
\usepackage{tikz}
\usetikzlibrary{trees,intersections,arrows}
\usepackage{url}
\usepackage{xspace}
\usepackage{shortcuts}

\ifCLASSOPTIONcompsoc
  \usepackage[nocompress]{cite}
\else
  \usepackage{cite}
\fi

\begin{document}

\title{Evaluating the Robustness of Trigger Set-Based Watermarks Embedded in
Deep Neural Networks}

\author{Suyoung~Lee,
        Wonho~Song,
        Suman~Jana,
        Meeyoung~Cha,
        and~Sooel~Son,~\IEEEmembership{Member,~IEEE}%
\IEEEcompsocitemizethanks{
\IEEEcompsocthanksitem Suyoung Lee, Wonho Song, Meeyoung Cha, and Sooel Son are
with School of Computing, KAIST, Daejeon 34141, South Korea.
\IEEEcompsocthanksitem Suman Jana is with Department of Computer Science,
Columbia University, New York, NY 10027, USA.}%
}

\IEEEtitleabstractindextext{%
\begin{abstract}
Trigger set-based watermarking schemes have gained emerging attention as they
provide a means to prove ownership for deep neural network model owners.
In this paper, we argue that state-of-the-art trigger set-based watermarking
algorithms do not achieve their designed goal of proving ownership.
We posit that this impaired capability stems from two common experimental
flaws that the existing research practice has committed when evaluating
the robustness of watermarking algorithms: (1) incomplete
adversarial evaluation and (2) overlooked adaptive attacks.
We conduct a comprehensive adversarial evaluation of \numwm representative
watermarking schemes against \numattack of the existing attacks and demonstrate
that each of these watermarking schemes lacks robustness against at least two
non-adaptive attacks.
We also propose novel adaptive attacks that harness the adversary's knowledge
of the underlying watermarking algorithm of a target model. We demonstrate that the
proposed attacks effectively break all of the \numwm watermarking schemes,
consequently allowing adversaries to obscure the ownership of any watermarked
model.
We encourage follow-up studies to consider our guidelines when evaluating the
robustness of their watermarking schemes via conducting comprehensive
adversarial evaluation that includes our adaptive attacks to demonstrate a
meaningful upper bound of watermark robustness.

\end{abstract}

}

\maketitle

\section{Introduction} \label{s:intro}
\IEEEPARstart{T}{he} recent advent of deep neural networks (DNNs) has
accelerated the development and application of diverse DNN models across various
domains, including image search~\cite{gordo:eccv:2016, wan:mm:2014},
security~\cite{xu:ccs:2017, ye:ccs:2018}, and self-driving
vehicles~\cite{tian:icse:2018}.
As machine learning technology evolves, the structures of state-of-the-art DNN
models have become more complicated. This trend renders corporations with fewer
computational resources unable to train state-of-the-art DNN models from scratch.
For instance, the ImageNet~\cite{russakovsky:ijcv:2015} dataset holds 14M
images; training a high-performing DNN model such as a
ResNet-50~\cite{he:cvpr:2016}, which consists of over 25M parameters, takes up to
several weeks with a machine equipped a Tesla M40 GPU.
Moreover, it is difficult to obtain a large number of high-quality
training instances pertaining to privacy-sensitive information, thus rendering
it infeasible for corporations with limited data access to produce a superb model.

An adversary may attempt to steal such a superb model and host another service
that imitates the service provided by the original model.
This adversary poses a grave threat to the model owner, who has invested resources
and time to develop a high-performing model. DNN model theft thus infringes on
the intellectual property (IP) of the model owner and discloses the owner's
business secrets.
Accordingly, corporations seek a mechanism that proves the ownership of their
DNN models to protect their IPs and business secrets.

Previous studies have proposed novel methods that validate
ownership of a given DNN model, thus protecting the owner's IP.
Similar to watermarking algorithms devised to protect the IP of multimedia
content, such as images and videos~\cite{langelaar:spm:2000, swanson:ieee:1998},
previous studies have proposed new ways of embedding watermarks into a given DNN
model as well as algorithms that verify ownership~\cite{uchida:icmr:2017,
adi:usec:2018, guo:iccad:2018, zhang:asiaccs:2018, chen:icmr:2019,
fan:nips:2019, li:acsac:2019, merrer:nca:2019, namba:asiaccs:2019,
rouhani:asplos:2019, zhang:aaai:2020, jia:usec:2021}.
The proposed watermarking algorithms are categorized into two types based on
their methods of embedding watermarks: feature-based and trigger set-based methods.

Feature-based schemes~\cite{uchida:icmr:2017, chen:icmr:2019,
rouhani:asplos:2019} require white-box access to a model's internal weight parameters.
On the other hand, trigger set-based watermarking methods~\cite{adi:usec:2018,
zhang:asiaccs:2018, guo:iccad:2018, fan:nips:2019, li:acsac:2019, jia:usec:2021,
merrer:nca:2019, namba:asiaccs:2019, zhang:aaai:2020} have gained attention
due to their comparative merits of requiring black-box access for ownership verification.
Trigger set-based schemes harness the common query interface of a suspect model.
Specifically, these watermarking methods leverage carefully created images, called
\emph{key images}. A model owner assigns an
arbitrary label, called a \emph{target label}, to the key images and
generates a \emph{trigger set} that consists of an arbitrary number of
key image and target label pairs. The owner then trains a model on this
trigger set as well as on the normal training data.
When verifying ownership, the model owner queries the model in doubt with the
key images and checks whether the model returns the target label; this enables
the owner to verify the ownership by using remote queries.
Previous trigger set-based methods~\cite{adi:usec:2018, zhang:asiaccs:2018,
guo:iccad:2018, fan:nips:2019, li:acsac:2019, merrer:nca:2019,
namba:asiaccs:2019, zhang:aaai:2020} have in common that they use key images
and target labels but differ in how they generate key images or select target
labels.

\noindent\textbf{Contributions.}
In this paper, we argue that trigger set-based watermarking methods today~\cite{
adi:usec:2018, zhang:asiaccs:2018, guo:iccad:2018,  fan:nips:2019,
jia:usec:2021, li:acsac:2019,merrer:nca:2019, namba:asiaccs:2019,
zhang:aaai:2020} do not achieve their goal of enabling model owners to prove
their ownership of watermarked DNN models.
We believe that previous studies have not evaluated the robustness of their
watermarking algorithms to the fullest extent, thereby failing to demonstrate
their readiness for real-world deployment.

There exist two different strategies for evaluating the robustness of
a DNN model: (1) proving a theoretical lower bound with
approximation~\cite{bastani:nips:2016, huang:cav:2017} and (2) demonstrating an
upper bound via adversarial evaluation with strong attacks.
Previous watermarking studies~\cite{adi:usec:2018, zhang:asiaccs:2018,
guo:iccad:2018, fan:nips:2019, li:acsac:2019, merrer:nca:2019, jia:usec:2021,
namba:asiaccs:2019, zhang:aaai:2020} have taken the latter approach,
demonstrating their robustness against selected attacks.
However, we observed two common
flaws in previous studies when evaluating the robustness of their trigger
set-based watermarking algorithms: (1) performing incomplete adversarial
evaluation and (2) overlooking an adaptive adversary.

\noindent\textbf{Incomplete adversarial evaluation.}
Because various attacks have been introduced across diverse studies
in various contexts, we first consolidate and reorganize six existing attacks.
We then categorize the existing attacks into two types, each of which
corresponds to either of the adversary's two strategies: (1) claiming ownership
by the adversary or (2) obscuring the owner's ownership.

We observed that no previous watermarking studies have considered the complete
set of the existing strong attacks in their adversarial evaluation; previous studies have not
demonstrated their robustness against at least one critical attack.
Furthermore, we contend that adversarial evaluation of attacks employing
the adversary's first strategy (claiming ownership by the adversary) is unnecessary.
This strategy permits for a target model to contain watermarks from its original owner
as well as the adversary, which demands additional proof to prove the adversary's
ownership (\S\ref{ss:shortcomings}). Therefore, there is no motive for the adversary to
employ this strategy alone unless she combines the two aforementioned strategies
by  obscuring the original watermarks and then injecting her own watermarks.

To this end, we perform our own adversarial evaluation against \numwm of
the representative trigger set-based watermarking schemes  while taking into
account the aforementioned problems. We demonstrate that
they are weak against at least two of these attacks.
In particular, all of the 11 evaluated schemes were vulnerable to
evasion~\cite{meng:ccs:2017, li:acsac:2019, namba:asiaccs:2019} and ownership
piracy attacks~\cite{adi:usec:2018}.

\noindent\textbf{Overlooked adaptive adversary.}
Previous studies focused on evaluating their watermark robustness against selected
existing attacks. Meanwhile, a vast volume of recent research on establishing the
robustness of DNN models has considered adaptive adversaries~\cite{
quiring:usec:2020,chen:usec:2020, shan:ccs:2020, carlini:aisec:2017, jia:ccs:2019,
cao:ndss:2021}.

To this end, we propose three novel attacks that a strong adaptive adversary is able to
conduct. Under the assumption that this adversary knows the underlying watermarking
algorithm of a target model, we demonstrate that the proposed adaptive attacks
effectively break all existing watermarking schemes, enabling the adversary to
obscure the ownership of a target model, regardless of its underlying
watermarking scheme.
Therefore, our proposed attacks contribute to demonstrating a new upper bound of
watermark robustness.

Overall, our experimental results demonstrate that trigger set-based
watermarking schemes today are far from ready for real-world deployment. We
recommend that future research evaluate their watermarking methods against at least all
existing strong attacks, including our adaptive attacks, and consider our guidelines
when demonstrating their watermark robustness via adversarial evaluation (\S\ref{sec:lessons}).

To enable follow-on research to evaluate its watermarking schemes, all of our
attack algorithms and their implementation will be available at \url{
https://github.com/WSP-LAB/wm-eval-zoo}.

\section{Background}

\subsection{DNN Ownership Verification}
Since Uchida~\etal~\cite{uchida:icmr:2017} proposed the first approach
to embedding watermarks into neural networks, various watermarking techniques
have been proposed. In terms of their watermark embedding methodology, these
watermarking techniques have been categorized into two types: trigger set-based
and feature-based methods.
Trigger set-based methods utilize additional training samples as watermarks
for DNNs~\cite{adi:usec:2018, guo:iccad:2018,
zhang:asiaccs:2018, fan:nips:2019, li:acsac:2019, merrer:nca:2019,
namba:asiaccs:2019, zhang:aaai:2020, jia:usec:2021}.
Feature-based methods embed watermarks by modifying model
features~\cite{uchida:icmr:2017, chen:icmr:2019, rouhani:asplos:2019}.

Zhang~\etal~\cite{zhang:asiaccs:2018} proposed a representative trigger set-based
method. They trained a model to learn predefined key pairs, each consisting
of a key image and its target label.
Specifically, they assigned a \emph{false} label with respect
to the ground-truth function to the key image.
The gist of their approach is that a model without the watermark is highly likely
to emit a ground-truth label rather than the predefined false label for a given
key image. Therefore, the owner can prove the ownership afterward by querying the
model with the key images and checking whether the model outputs the predefined
false label. In this scheme, the key images and their predefined false labels
become a \emph{trigger set}.

Other trigger set-based watermarking techniques employ more or less similar
approaches, but Adi~\etal~\cite{adi:usec:2018} further integrated this scheme
with cryptographic primitives to secure embedded watermarks. Recently,
Jia~\etal~\cite{jia:usec:2021} proposed to train a model in the direction of
tightly coupling the trigger set with a regular training set so that the trained
model becomes robust against model stealing attacks.

\subsection{Target Watermark Schemes}
\label{ss:targetwms}
Our goal is to evaluate the robustness of state-of-the-art trigger
set-based watermark schemes.
Thus, we chose \numwm representative watermarking algorithms, published at top
venues over the past five years~\cite{zhang:asiaccs:2018, guo:iccad:2018, adi:usec:2018,
merrer:nca:2019, fan:nips:2019, li:acsac:2019, namba:asiaccs:2019,
rouhani:asplos:2019, jia:usec:2021}.
They share a common scheme that uses trigger sets for verifying ownership.

\begin{algorithm}[htb] {
  \footnotesize
  \DontPrintSemicolon
  \SetNoFillComment
  \SetKwSty{algokeywordsty}
  \SetFuncSty{algofuncsty}
  \SetDataSty{algodatasty}
  \SetArgSty{algoargsty}
  \SetKwInOut{Input}{Input}
  \SetKwInOut{Output}{Output}
  \SetKwFunction{embedWM}{\texttt{EmbedWatermark}}
  \SetKwFunction{genKeyImg}{\texttt{GenerateKeyImgs}}
  \SetKwFunction{genTargetLabel}{\texttt{AssignTargetLabels}}
  \SetKwFunction{train}{\texttt{TrainModel}}
  \SetKwProg{Fn}{function}{}{}

  \Input{
    A regular training set (\trainset).
    \\ A set of source images (\srcimg).
  }

  \Output{
    A watermarked model (\wmmodel).
  }

  \Fn{\embedWM{\trainset, \srcimg}}{
    \keyimg $\gets$ \genKeyImg{\srcimg} \; \label{algo:line:genimg}
    \targetlabel $\gets$ \genTargetLabel{\keyimg} \; \label{algo:line:genlabel}
    \triggerset $\gets$ (\keyimg, \targetlabel) \; \label{algo:line:trigger}
    \wmmodel $\gets$ \train{\trainset, \triggerset} \; \label{algo:line:train}
    \Return{\wmmodel}
  }
}
  \caption{Embedding a trigger set into a DNN.}
  \label{algo:watermark}
\end{algorithm}

Algorithm~\ref{algo:watermark} summarizes how a trigger set-based watermark
algorithm embeds the ownership proof of an owner \emph{O} into a DNN
model. \emph{O} provides a training set \trainset  and a set of source
images \srcimg to the \texttt{EmbedWatermark} function.
Given \srcimg, \texttt{GenerateKeyImgs}  generates  a set of key images
\keyimg (Line~\ref{algo:line:genimg}).
Note that these key images are intentionally designed to have a different
underlying distribution than that of \trainset. Owing to the
over-parameterization of DNN models, they are capable of intentionally learning
key images along with \trainset~\cite{zhang:iclr:2017, choromanska:aistats:2015}.
The \texttt{AssignKeyLabels} function assigns a target label \targetlabel
to each key image (Line~\ref{algo:line:genlabel}). We call generated key images
together with their assigned target labels as a \emph{trigger set} \triggerset.
Finally, the \texttt{TrainModel} function trains a model with both \trainset and
\triggerset to embed watermarks (Line \ref{algo:line:train}).
This step is analogous to backdoor attacks~\cite{gu:arxiv:2019, chen:arxiv:2017}
\emph{per se} but different in that this step is used to claim ownership of a
DNN model, instead of emplacing backdoors.

When \emph{O} claims her ownership, she conducts the following verification
phase: \emph{O} queries a model in doubt with the key images. If the model is
indeed the owner's genuine model, the model will output the predefined target
labels trained in the training phase.
In Supplemental Material~1~\cite{lee:tdsc-sup:2022}, we describe each of the \numwm selected
watermarking algorithms. Throughout the paper, we denote each algorithm as
follows:
\noindent
\wmcontent~\cite{zhang:asiaccs:2018},
\wmnoise~\cite{zhang:asiaccs:2018}, \wmunrelated~\cite{zhang:asiaccs:2018},
\wmmark~\cite{guo:iccad:2018}, \wmabstract~\cite{adi:usec:2018},
\wmadv~\cite{merrer:nca:2019}, \wmpassport~\cite{fan:nips:2019},
\wmencoder~\cite{li:acsac:2019}, \wmexp~\cite{namba:asiaccs:2019},
DeepSigns~\cite{rouhani:asplos:2019}, and \wmentangled~\cite{jia:usec:2021}.

\section{Adversary Model}
\label{s:adversary}

We introduce an attack scenario in which an adversary infringes on the IP of a
model owner with an exfiltrated DNN model, along with the notations that we use
throughout the paper. We then describe the prior knowledge of an adversary regarding
the exfiltrated model.

\subsection{Attack Scenario}
\label{ss:attack-scenario}

We assume two parties in the attack scenario: a model owner \owner and an
adversary \adv. \owner embeds watermarks into a neural network
model \orgmodel by training \orgmodel with a trigger set, thus producing the
watermarked model \wmmodel. \owner then hosts a service by leveraging \wmmodel.
On the other hand, \adv decides to steal \wmmodel because training a precise
model from scratch requires a lot of computational resources as well as training
instances. For instance, \adv can steal \wmmodel by compromising \owner's
machine learning service server or getting help from an insider. Enumerating
the feasible ways of \adv obtaining \wmmodel is beyond the scope of this paper.

After stealing \wmmodel, \adv hosts a similar service as \owner using a model
\advmodel derived from \wmmodel.
Note that the end goal of \adv is to either (1) obscure \owner's ownership of
\advmodel or (2) claim the ownership of \advmodel. Therefore, \adv may have built
\advmodel by transforming \wmmodel to achieve these goals. That is, \wmmodel and
\advmodel are not necessarily the same.
We further elaborate on attack scenarios with these goals in
\S\ref{ss:attack-overview}.

Finally, once \owner suspects that \advmodel is derived from \wmmodel, \owner will
attempt to prove their ownership of \advmodel.
However, if \owner watermarked \wmmodel with a feature-based scheme, \owner must
have white-box access to \advmodel to verify the ownership.
Considering that \adv certainly wants to hide the true ownership of \advmodel,
\adv will not provide white-box access to \advmodel unless \advmodel is under
litigation. Thus, in this paper, we focus on trigger set-based watermark
schemes, which only require black-box access for ownership verification.

\subsection{Adversarial Knowledge}
\label{ss:adv-knowledge}

We assume two adversaries according to their adversarial knowledge:
(1) a non-adaptive adversary and (2) an adaptive adversary.
A non-adaptive adversary knows that the stolen target model \wmmodel has been
watermarked but does not know which specific watermarking algorithm was used.
On the other hand, an adaptive adversary knows the exact watermarking
algorithm that \owner harnessed to protect the model among various trigger
set-based methods. Specifically, the adaptive adversary only knows the internal
working of \texttt{GenerateKeyImgs} in Algorithm~\ref{algo:watermark}.
She does not know the source images (\srcimg) for \texttt{GenerateKeyImgs}.
She also has no access to the original trigger set (\triggerset) as well as the
training dataset (\trainset).

Note that both adversaries share the same knowledge except about the
watermarking algorithm.
As both adversaries stole \wmmodel from \owner, they can observe the model
inputs, outputs, and structure.
Additionally, we assume that they have access to 50\% of a testing set, which is
required to launch attacks against \wmmodel.
Note that this data accessible by the adversaries is completely disjointed from
the original training set, assuming the least privilege granted to them.
Previous studies~\cite{li:acsac:2019, adi:usec:2018, zhang:asiaccs:2018,
rouhani:asplos:2019} assume similar capabilities for the adversary to conduct
different attacks.
We further considered adversaries who have access to fewer data in Supplemental
Material~4~\cite{lee:tdsc-sup:2022}.

\section{Motivation} \label{s:problems}

We argue that today's evaluation practice of demonstrating watermark robustness exhibits
two common shortcomings:  incomplete adversarial evaluation (\S\ref{ss:problem-1}) and
overlooked adaptive attacks (\S\ref{ss:problem-2}).

\subsection{Incomplete Adversarial Evaluation}
\label{ss:problem-1}

\newcolumntype{Y}{>{\centering\arraybackslash}X}

\begin{table}
  \begin{threeparttable}
    \footnotesize
    \centering
    \renewcommand{\arraystretch}{1.1}
    \caption{Summary of adversarial evaluations performed by previous studies.}
    \label{tab:attack}
    \begin{tabularx}{\columnwidth}{lYYYYYYYYYYY}
      \toprule
        \textbf{Attack} &
        \rotate{\wmcontent} &
        \rotate{\wmnoise} &
        \rotate{\wmunrelated} &
        \rotate{\wmmark} &
        \rotate{\wmabstract} &
        \rotate{\wmadv} &
        \rotate{\wmpassport} &
        \rotate{\wmencoder} &
        \rotate{\wmexp} &
        \rotate{DeepSigns} &
        \rotate{\wmentangled} \\

      \midrule

        Fine-tuning &
        \cmark &  %
        \cmark &  %
        \cmark &  %
        \xmark &  %
        \cmark &  %
        \xmark &  %
        \cmark &  %
        \cmark &  %
        \xmark &  %
        \cmark &  %
        \cmark \\ %

        \rowcolor{lightgray}
        Model Stealing &
        \xmark &  %
        \xmark &  %
        \xmark &  %
        \xmark &  %
        \xmark &  %
        \xmark &  %
        \xmark &  %
        \xmark &  %
        \xmark &  %
        \xmark &  %
        \cmark \\ %

        Parameter Pruning &
        \cmark &  %
        \cmark &  %
        \cmark &  %
        \xmark &  %
        \xmark &  %
        \cmark &  %
        \cmark &  %
        \xmark &  %
        \cmark &  %
        \cmark &  %
        \cmark \\ %

        \rowcolor{lightgray}
        Evasion &
        \xmark &  %
        \xmark &  %
        \xmark &  %
        \xmark &  %
        \xmark &  %
        \xmark &  %
        \xmark &  %
        \cmark &  %
        \cmark &  %
        \xmark &  %
        \cmark \\ %

      \midrule
        Ownership Piracy &
        \xmark &  %
        \xmark &  %
        \xmark &  %
        \xmark &  %
        \cmark &  %
        \cmark &  %
        \xmark &  %
        \xmark &  %
        \xmark &  %
        \cmark &  %
        \cmark \\ %

        \rowcolor{lightgray}
        Ambiguity &
        \cmark &  %
        \cmark &  %
        \cmark &  %
        \xmark &  %
        \xmark &  %
        \xmark &  %
        \cmark &  %
        \xmark &  %
        \xmark &  %
        \xmark &  %
        \xmark \\ %

      \midrule
        \textbf{\# of Evaluated Attacks} &
        3 &  %
        3 &  %
        3 &  %
        0 &  %
        2 &  %
        2 &  %
        3 &  %
        2 &  %
        2 &  %
        3 &  %
        5 \\ %
      \bottomrule
    \end{tabularx}
  \end{threeparttable}
\end{table}

We observe that previous studies on trigger set-based watermarks have evaluated
the robustness of their methods using arbitrary choices of the existing attacks, thus
demonstrating an upper bound on their robustness only to the selected attacks.
Due to the nature of adversarial evaluation, the existence of one effective attack
denotes the failure to protect the IP of \owner, effectively breaking a target
watermarking scheme.
Therefore, it is paramount to account for all existing attacks to demonstrate
meaningful robustness.

Table~\ref{tab:attack} summarizes the evaluations performed by the previous
watermark research in terms of applicable existing attacks.
Note from the table that no previous studies evaluated their approaches
against a complete set of attacks.
Among the six attacks, 10 out of \numwm prior watermark studies only considered
at most three attacks and missed other attacks in their evaluations.
Moreover, model stealing attacks have never been evaluated in any previous
studies.

We emphasize that all six attacks examined herein have existed since each watermarking
algorithm was first proposed.
In other words, ever since each watermarking algorithm was first proposed, their
robustness against several existing state-of-the-art attacks has remained
unexplored.
Therefore, it is still questionable whether state-of-the-art watermarking
algorithms can successfully work as a defense mechanism against various
real-world threats.

Furthermore, incomplete adversarial evaluation becomes problematic when
comparing the robustness of different watermarking algorithms.
Because the previous studies evaluated watermarking algorithms against
arbitrarily chosen attacks, they have failed to demonstrate which algorithms
are more robust than others in general.
Even though one algorithm is robust against a given attack, it can be broken by
another attack against which other algorithms are known to be secure.
We believe that this incomplete evaluation practice stems from the lack of prior
systematic studies that enumerate all the applicable attacks. Thus, in this
paper, we summarize these attacks (\S\ref{ss:attack-overview}).

\subsection{Overlooked Adaptive Attacks}
\label{ss:problem-2}

A vast volume of recent research on securing machine learning models has striven
to demonstrate a meaningful upper bound of its
robustness~\cite{quiring:usec:2020, chen:usec:2020, shan:ccs:2020,
carlini:aisec:2017, jia:ccs:2019, cao:ndss:2021}.
To this end, they have focused on strong adaptive adversaries who know the
adopted defense algorithms for securing the model.
Nevertheless, the previous studies on watermarking algorithms have not yet taken
into account adaptive attacks in their adversarial evaluation.
Therefore, to challenge the robustness of watermarking algorithms to the fullest
extent, we propose novel adaptive attacks in the context of DNN watermarking.

Note that the existing attacks in Table~\ref{tab:attack} are non-adaptive
attacks.
In addition to these attacks, we consider adaptive attacks against \wmmodel.
The adaptive adversary mounts the same attacks as non-adaptive adversaries.
She leverages her prior knowledge of the underlying watermarking algorithm
and adapts these attacks, thus mounting strong attacks.

\section{Attack Algorithms}
\label{s:attack}

We now introduce state-of-the-art attacks that  non-adaptive and adaptive
adversaries (\S\ref{s:adversary}) can conduct.
We consolidate six of the existing attacks spread across various studies in the
literature and systematically categorize them from the perspective of the goal
that the adversaries aim to achieve (\S\ref{ss:attack-overview}).
We then briefly describe each existing attack (\S\ref{ss:attack-scenario-1}--\S\ref{ss:attack-scenario-2}).
Finally, we present novel attacks that the adaptive adversary is able to conduct
via leveraging the knowledge of a target watermarking algorithm (\S\ref{ss:adaptive}).

\subsection{Attack Overview}
\label{ss:attack-overview}

An adversary \adv can devise two different scenarios to conceal the fact that
\adv stole \wmmodel from \owner; \adv can decide to either obscure \owner's
ownership or claim her ownership.

\noindent\textbf{Obscuring \owner's ownership.}
The goal of \adv in this scenario is to thwart \owner's ownership
verification by modifying \wmmodel, such as by training a counterfeit model
or detecting key images. As \owner fails to verify their ownership in this scenario,
\adv can successfully obscure \owner's ownership and insist that \wmmodel is not
watermarked. To achieve this goal, \adv can launch fine-tuning, model
stealing, evasion, or parameter pruning attacks.

\noindent\textbf{Claiming ownership by \adv.}
Another scenario that \adv can consider is to claim the ownership of \wmmodel by
implanting a new trigger set into \wmmodel or generating a set of fake key
images that can trigger the target labels.
Note that \adv does not aim to damage \owner's ownership and \owner's watermark
may persist. Therefore, both \owner and \adv can claim the ownership based on the
respective trigger set, which results in conflicting ownership arguments.
Since it is infeasible to decide which one is fraudulently claiming ownership
solely based on their key images and target labels, previous
studies~\cite{merrer:nca:2019, rouhani:asplos:2019, adi:usec:2018,
fan:nips:2019, zhang:asiaccs:2018, jia:usec:2021} have considered this to be a
plausible strategy.
To realize this scenario, \adv is able to conduct one of the following two
attacks: ownership piracy or ambiguity attacks.

\subsection{Obscuring \owner's Ownership}
\label{ss:attack-scenario-1}

\noindent\textbf{Fine-tuning attack.}
To remove the original watermark, \adv can fine-tune \wmmodel with a new
training set~\cite{zhang:asiaccs:2018, adi:usec:2018, fan:nips:2019,
li:acsac:2019, rouhani:asplos:2019, chen:asiaccs:2021, jia:usec:2021}.
Specifically, \adv trains \wmmodel with a new small set that shares an
underlying distribution with the original training set, thus preventing \wmmodel
from losing its original functionality. At the same time, \adv does not
include any data that are distant from the underlying distribution in the new
training set in the expectation that \wmmodel will forget \owner's key images.

\noindent\textbf{Model stealing attack.}
\adv in model stealing attacks~\cite{tramer:usec:2016, orekondy:cvpr:2019,
jia:usec:2021} aims to copy the functionality of \wmmodel into a new model,
except for the capability of remembering the trigger set. To this end, \adv
labels arbitrary images by querying \wmmodel. Using the constructed training
set, \adv trains a model from scratch. The new model may forget \owner's key
images because the distribution represented by the arbitrary images is highly
likely not to include \owner's trigger set.

\noindent\textbf{Parameter pruning attack.}
As an attempt to make \wmmodel forget a trained trigger set, \adv in parameter
pruning attack scenarios~\cite{zhang:asiaccs:2018, merrer:nca:2019,
fan:nips:2019, namba:asiaccs:2019, rouhani:asplos:2019, jia:usec:2021} prunes certain
parameters of \wmmodel.  The original goal of model pruning is to reduce the
number of redundant parameters in DNNs. However, recall that model watermarking
is possible due to the over-parameterization of DNNs. \adv expects \wmmodel to
lose the capability of remembering the key images after the pruning of some
trained parameters, thus causing \owner's ownership claim to fail.

\noindent\textbf{Evasion attack.}
\adv conducting evasion attacks~\cite{meng:ccs:2017, namba:asiaccs:2019,
jia:usec:2021, li:acsac:2019} may attempt to detect key images on the fly when
\owner queries \wmmodel.  Recall that key images do not belong to the underlying
distribution of regular images. Thus, \adv can distinguish key images by
checking the distribution of a given image. Once \adv finds a suspicious image,
she can evade the verification process by returning a random label.

\subsection{Claiming Ownership by \adv}
\label{ss:attack-scenario-2}

\noindent\textbf{Ownership piracy attack.}
In ownership piracy attacks~\cite{adi:usec:2018, rouhani:asplos:2019,
jia:usec:2021, merrer:nca:2019}, \adv attempts to implant her own new trigger
set into \wmmodel to claim the ownership.
Specifically, \adv prepares a new trigger set that is different from the
original and then retrains \wmmodel with the new trigger set. After training,
\advmodel will classify \adv's key images as their target labels, and \adv
can fraudulently claim the ownership of \advmodel, which leads to conflicting
ownership arguments.

\noindent\textbf{Ambiguity attack.}
To claim ownership, in an ambiguity attack scenario~\cite{zhang:asiaccs:2018,
fan:nips:2019, fredrikson:ccs:2015}, \adv generates a set of counterfeit key
images that can trigger the target labels.
Similar to model inversion attacks~\cite{fredrikson:ccs:2015}, \adv gradually
updates regular images by leveraging gradient descent so that \wmmodel
classifies the updated images as their predefined labels.
The core difference of this attack compared to ownership piracy attacks is that
the adversary in this scenario does not modify \wmmodel but creates counterfeit
key images by leveraging \wmmodel.

Assume a scenario where \adv launches an ambiguity attack against \wmmodel
trained on CIFAR-10 and watermarked using \wmcontent.
\adv can add quasi-imperceptible perturbations to ``apple'' images taken from
CIFAR-100 such that \wmmodel classifies each image as an ``airplane.'' In this
scenario, \adv can verify the ownership based on \wmunrelated using the perturbed
images as key images.

\subsubsection{Shortcomings of Evaluation}
\label{ss:shortcomings}

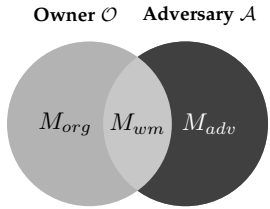
\begin{figure}[!t]
\centering
  \resizebox{.4\columnwidth}{!}{%
\begin{tikzpicture}
  \def\own{(0,0) circle (1.2)}
  \def\adversary{(1.4,0) circle (1.2)}
  \fill[gray] \adversary;
  \begin{scope}[blend group = screen]
    \fill[gray!60] \own;
    \fill[darkgray] \adversary;
  \end{scope}

  \node [font=\normalsize] at (-0.35,0) {\orgmodel} ;
  \node [font=\normalsize] at (0.7,0) {\wmmodel} ;
  \node [font=\normalsize, text=white] at (1.75,0) {\advmodel} ;
  \node [font=\footnotesize] at (-0.2,1.6) {\textbf{Owner \owner}} ;
  \node [font=\footnotesize] at (1.6,1.57) {\textbf{Adversary \adv}} ;
\end{tikzpicture}
}
\caption{The difference between models accessible to \adv who aims to claim
her ownership and \owner.}
\vspace{-1.2em}
\label{fig:diff}
\end{figure}

Recall from \S\ref{ss:attack-overview} that ownership piracy and ambiguity
attacks inevitably cause a stalemate between \adv and \owner with conflicting
ownership arguments based on their respective watermarks. In this regard,
previous studies~\cite{zhang:asiaccs:2018, fan:nips:2019, adi:usec:2018,
jia:usec:2021, merrer:nca:2019, rouhani:asplos:2019} have demonstrated the
degree to which their watermarking algorithms can withstand these attacks.
However, we claim that there exists a straightforward solution to manifest the
true owner in these attack scenarios; thus, their evaluation should have been
performed assuming a different scenario.

We note that there exists a clear difference between the capabilities of \owner
and \adv, as shown in Figure~\ref{fig:diff}.
Because \adv steals the model after
\owner watermarks \orgmodel, \adv cannot access \orgmodel which does not have
any watermarks.
Accordingly, in court, a judge may request that both \owner and \adv provide
a functional model without any watermarks. Then, \owner can prove the ownership by
providing \orgmodel, which \adv cannot provide.
\adv will lose this ownership dispute game due to the inability to present a functional
model that remembers none of the key images and achieves a test accuracy
comparable to \advmodel at the same time.

We emphasize that \adv in ownership piracy or ambiguity attack scenarios does
not possess the aforementioned functional model without key images.
One may argue that \adv can present this model to the court by conducting an
attack that removes \owner's watermark.
However, if \adv was able to remove \owner's watermark, the ownership dispute
would not have occurred in the first place because the attacker would have used
the watermark-removed model for hosting the service.

This verification leveraging the adversary's inability of presenting a watermark-free
model is analogous to that in the traditional image and video watermarking
research~\cite{craver:jsac:1998, adelsbach:drm:2003}.
To prevent the threat of an adversary claiming ownership by means of blending her
watermarks on top of the owner’s watermarked image, it is common in the
verification to ask the adversary to present the original image without any watermarks.

Therefore, we propose the following more plausible scenario in which \adv
aims to obscure \emph{O}'s ownership and claim her ownership at the same time.
To achieve both goals together, we insist that \adv should first mount an
attack that removes \owner's watermark and then launch attacks devised to claim
\adv's ownership against the watermark-removed model, thus constructing a model
that only remembers \adv's trigger set.
Unfortunately, no previous studies have considered these attacks
together. On the contrary, we considered this new scenario by performing
ownership piracy and ambiguity attacks against target models after removing \owner's
watermark (\S\ref{ss:eval-claiming}).

\subsection{Adaptive Attacks}
\label{ss:adaptive}

We argue that the robustness of watermarked models should not be undermined
by the adversary's prior knowledge of target watermarking algorithms.
Considering that any insiders are able to leak the algorithms, solely depending on
the security by obscurity is not a desirable goal that follow-up watermarking
studies should pursue. Carlini~\etal have also emphasized the necessity of evaluations
against adaptive attacks for demonstrating adversarial robustness~\cite{carlini-eval:arxiv:2019}.

To this end, we propose novel adaptive attacks in which the adversary can adapt their
attacks to a given watermarking scheme.
In adaptive attacks, the adversary aims to obscure \owner's ownership by
modifying \wmmodel to remove \owner's trigger set.
For this, the adaptive adversary removes \owner's trigger set by employing
the same fine-tuning, model stealing, and pruning attacks (\S\ref{ss:attack-scenario-1}).
The key difference is that this adversary engineers a new trigger set
that plays a role similar to \owner's trigger set against \wmmodel and leverages
this new trigger set when launching the aforementioned three watermark
removal attacks.
In the following, we explain how the adversary can adaptively create the new trigger set based on \wmmodel's watermarking algorithm.

We propose a general framework that the adaptive adversary leverages to create
a new trigger set.
Since the adversary seeks to generate new key images that serve as \owner's key
images, the new key images should have an underlying distribution similar to that of
the original key images. At the same time, the new key images should be able to
trigger attacker-specified target labels.
To achieve these two goals, we propose to train an autoencoder such that (1) the output
images have a distribution similar to images that the watermarking scheme of \wmmodel
generates and (2) \wmmodel classifies each output image as a target label.
Note that the adaptive adversary can train such an autoencoder by leveraging her
knowledge about the target watermarking scheme and white-box access to the
stolen target model.
Specifically, given a source image $x$ and a target label $y_t$, the adversary
trains the autoencoder to minimize the following loss function.
\begin{equation}
\small
  \begin{aligned}
  \label{eq:loss}
    x' &= AutoEncoder(x) \\
    L(x, y_t)  &= L_{ae}(x, x') + \lambda \cdot L_f(y_t, f(x'))
  \end{aligned}
\end{equation}

In Equation~\ref{eq:loss}, the loss function has two terms: $L_{ae}$
and $L_f$. These terms are designed to achieve the autoencoder's two training
objectives, respectively.
$L_{ae}$ refers to a relationship between the input and output images
that the adversary can adaptively define based on a target watermarking scheme, and
$L_{f}$ refers to the classification error of a target model.

To perform strong attacks, it is important to choose well-suited source images $x$
and a loss function $L_{ae}$ so that the autoencoder is able to learn how a target
watermarking scheme performs the \texttt{GenerateKeyImgs} function in
Algorithm~\ref{algo:watermark} with high fidelity.
For instance, consider \wmabstract~\cite{adi:usec:2018} as a target
watermarking scheme.
In this case, the adversary can use arbitrary abstract images collected from the
Internet as source images and choose the mean squared error loss function as
$L_{ae}$ so that the output images $x'$ become abstract images that can trigger
target labels when given to \wmmodel.
We describe the source images and loss functions that we chose to model each of
our target watermarking schemes in Supplemental Material~2~\cite{lee:tdsc-sup:2022}.
Note that we have devised fine-tuning, model stealing, and parameter pruning
adaptive attacks for each watermarking scheme, yielding 30 attack variants
(3 attacks $\times$ \numwm schemes).

Besides the source images $x$ and the loss function $L_{ae}$, the adaptive
adversary also needs to specify the target label $y_t$ to train this autoencoder
but has no prior knowledge about the target labels of the original key images.
Therefore, the adversary repeatedly trains this autoencoder for each class
while assuming the current class as a target label. Then, the adversary collects
trigger set pairs $(x', y_t)$ from all trained autoencoders and leverages all the
collected pairs when initiating the watermark removal attacks.
The adversary expects these trigger set pairs to effectively contribute to
removing the original trigger set of a target model.

\section{Implementation}
\label{s:impl}

We implemented the target watermarking algorithms and attacks using TensorFlow
2.7.0. However, publicly available code for \wmpassport and \wmentangled are
written in PyTorch 1.10.1 and TensorFlow 1.14.0, respectively. Since it requires
a huge engineering effort to migrate them to TensorFlow 2.7.0, we used the
corresponding frameworks to implement the attacks targeting these two schemes.
The remaining nine target algorithms were implemented by referring to their
papers and code if available.

\section{Evaluation}
In this section, we evaluate the robustness of the \numwm trigger set-based
watermarks. We first explain the datasets and DNN models that we used
(\S\ref{ss:setup}) and demonstrate how we successfully implanted watermarks
into the DNN models using the target watermark schemes in our experimental
settings (\S\ref{ss:repro}).
We then conduct the adversarial evaluation of each attack that we have
discussed so far (\S\ref{ss:eval-finetuning}--\S\ref{ss:eval-ambiguity}).

\subsection{Datasets and Target Models}
\label{ss:setup}

\noindent\textbf{Dataset.}
We use the MNIST, GTSRB, CIFAR-10, TinyImageNet, and CIFAR-100 datasets.
All the prior studies have only evaluated their algorithms using at most four
datasets.
We use these five widely adopted datasets of various sizes for extensive
evaluation.

\noindent\textbf{DNN models.}
For MNIST and TinyImageNet, we prepared LeNet-5 models~\cite{lecun:mnist:1998}
and EfficientNetV2S models~\cite{tan:icml:2021}. For the remaining datasets, we
implemented ResNet-56 models~\cite{he:cvpr:2016}. However, we employed ResNet-18
for all five datasets to evaluate \wmpassport and \wmentangled in the same setup
as provided by the authors (recall \S\ref{s:impl}).
Note that these models have been widely adopted in previous studies~\cite{
guo:iccad:2018, fan:nips:2019, adi:usec:2018, namba:asiaccs:2019, li:acsac:2019}.
\sooelnote{why three, not four?}
Since these three models show outstanding performance, they are highly likely to
be deployed in real-world cases, rendering them good target models for watermark
implantation.

\subsection{Embedding Watermarks into the DNN Models}
\label{ss:repro}

\newcolumntype{L}[1]{>{\raggedright\arraybackslash}m{#1}}
\newcolumntype{Y}[1]{>{\centering\arraybackslash}m{#1}}

\begin{table}
  \begin{threeparttable}
    \scriptsize
    \centering
    \renewcommand{\arraystretch}{1.15}
    \caption{Performance of the target models \wmmodel on four datasets: MNIST
    (MN), GTSRB (GT), CIFAR-10 (C10), TinyImageNet (TI), and CIFAR-100 (C100). Numbers in
    parentheses denote the degree to which test accuracy dropped compared to a model
    without watermarks.}
    \label{tab:repro}
    \setlength\tabcolsep{0.065cm}
    \begin{tabularx}{\columnwidth}{
        L{1.10cm}|
        Y{0.55cm}
        Y{0.55cm}
        Y{0.55cm}
        Y{0.55cm}
        Y{0.55cm}|
        Y{0.73cm}
        Y{0.73cm}
        Y{0.73cm}
        Y{0.73cm}
        Y{0.73cm}
      }
      \toprule
        &
        \multicolumn{5}{c}{\textbf{Trigger Set Recall (\%)}} &
        \multicolumn{5}{c}{\textbf{Test Acc. (\%)}} \\

      \cmidrule(lr){2-6}
      \cmidrule(lr){7-11}

        &
        \textbf{MN} &
        \textbf{GT} &
        \textbf{C10} &
        \textbf{TI} &
        \textbf{C100} &
        \textbf{MN} &
        \textbf{GT} &
        \textbf{C10} &
        \textbf{TI} &
        \textbf{C100} \\

      \midrule

        \multirow{2}{*}{Content} &
        \multirow{2}{*}{99.90} &  %
        \multirow{2}{*}{100}      &  %
        \multirow{2}{*}{100}   &  %
        \multirow{2}{*}{100}   &  %
        \multirow{2}{*}{100}   &  %
        98.85                  &  %
        94.75                  &  %
        93.09                  &  %
        77.16                  &  %
        71.71                  \\ %

        &
        &  %
        &  %
        &  %
        &  %
        &  %
        (-0.21) &  %
        (-0.28) &  %
        (0.12)  &  %
        (-0.74) &   %
        (-0.66) \\ %

        \multirow{2}{*}{Noise} &
        \multirow{2}{*}{100}   &  %
        \multirow{2}{*}{100}      &  %
        \multirow{2}{*}{100}   &  %
        \multirow{2}{*}{100}   &  %
        \multirow{2}{*}{100}   &  %
        99.04                  &  %
        94.89                  &  %
        93.20                  &  %
        78.55                  &  %
        72.77                  \\ %

        &
        &  %
        &  %
        &  %
        &  %
        &  %
        (-0.02)  &  %
        (-0.13)  &  %
        (0.23)   &  %
        (0.65)   & %
        (0.40)  \\ %

        \multirow{2}{*}{Unrelated} &
        \multirow{2}{*}{99.94}   &  %
        \multirow{2}{*}{100}      &  %
        \multirow{2}{*}{100}   &  %
        \multirow{2}{*}{100}   &  %
        \multirow{2}{*}{100}   &  %
        99.02                  &  %
        94.32                  &  %
        92.91                  &  %
        78.49                  &  %
        72.51                  \\ %

        &
        &  %
        &  %
        &  %
        &  %
        &  %
        (-0.04) &  %
        (-0.70)  &  %
        (-0.06) &  %
        (0.59)  &  %
        (0.14) \\ %

        \multirow{2}{*}{Mark} &
        \multirow{2}{*}{98.72} &  %
        \multirow{2}{*}{99.72}      &  %
        \multirow{2}{*}{99.64} &  %
        \multirow{2}{*}{99.90}   &  %
        \multirow{2}{*}{94.79} &  %
        98.94                  &  %
        96.56                  &  %
        92.42                  &  %
        73.96                  &  %
        70.85                  \\ %

        &
        &  %
        &  %
        &  %
        &  %
        &  %
        (-0.12) &  %
        (1.53)  &  %
        (-0.55) &  %
        (-3.94) &  %
        (-1.52) \\ %

        \multirow{2}{*}{Abstract} &
        \multirow{2}{*}{100} &  %
        \multirow{2}{*}{100}      &  %
        \multirow{2}{*}{100} &  %
        \multirow{2}{*}{100}   &  %
        \multirow{2}{*}{100}   &  %
        99.02                  &  %
        95.08                  &  %
        92.93                  &  %
        78.08                  &   %
        72.46                  \\ %

        &
        &  %
        &  %
        &  %
        &  %
        &  %
        (-0.04) &  %
        (0.06)  &  %
        (-0.04) &  %
        (0.18)  &  %
        (0.09)  \\ %

        \multirow{2}{*}{Adv} &
        \multirow{2}{*}{100}   &  %
        \multirow{2}{*}{100}      &  %
        \multirow{2}{*}{100}   &  %
        \multirow{2}{*}{99.00}   &  %
        \multirow{2}{*}{100}   &  %
        99.21                  &  %
        97.21                  &  %
        91.79                  &  %
        77.76                  &   %
        71.78                  \\ %

        &
        &  %
        &  %
        &  %
        &  %
        &  %
        (0.15) &  %
        (2.18)  &  %
        (-1.18) &  %
        (-0.14) &  %
        (-0.59) \\ %

        \multirow{2}{*}{Passport} &
        \multirow{2}{*}{84.00} &  %
        \multirow{2}{*}{100}   &  %
        \multirow{2}{*}{82.00} &  %
        \multirow{2}{*}{100}   &  %
        \multirow{2}{*}{93.00} &  %
        99.12                  &  %
        94.29                  &  %
        88.63                  &  %
        60.35                  &   %
        63.17                  \\ %

        &
        &  %
        &  %
        &  %
        &  %
        &  %
        (-0.23) &  %
        (0.95)  &  %
        (-2.72) &  %
        (-3.12) &  %
        (-4.87) \\ %

        \multirow{2}{*}{Encoder} &
        \multirow{2}{*}{100} &  %
        \multirow{2}{*}{96.94}      &  %
        \multirow{2}{*}{99.20} &  %
        \multirow{2}{*}{99.80}   &  %
        \multirow{2}{*}{99.20} &  %
        98.98                  &  %
        93.13                  &  %
        92.67                  &  %
        77.15                  &   %
        72.19                  \\ %

        &
        &  %
        &  %
        &  %
        &  %
        &  %
        (-0.08) &  %
        (-1.90)  &  %
        (-0.30) &  %
        (-0.75)    &  %
        (-0.18) \\ %

        \multirow{2}{*}{Exp} &
        \multirow{2}{*}{100} &  %
        \multirow{2}{*}{100}      &  %
        \multirow{2}{*}{100}   &  %
        \multirow{2}{*}{100}   &  %
        \multirow{2}{*}{100}   &  %
        99.07                  &  %
        94.54                  &  %
        92.62                  &  %
        77.30                  &   %
        71.53                  \\ %

        &
        &  %
        &  %
        &  %
        &  %
        &  %
        (0.01) &  %
        (-0.49)  &  %
        (-0.35) &  %
        (-0.60) &  %
        (-0.84) \\ %

        \multirow{2}{*}{DeepSigns} &
        \multirow{2}{*}{100}   &  %
        \multirow{2}{*}{100}      &  %
        \multirow{2}{*}{100}   &  %
        \multirow{2}{*}{100}   &  %
        \multirow{2}{*}{100}   &  %
        99.09                  &  %
        95.79                 &  %
        91.69                  &  %
        77.21                  &   %
        70.27                  \\ %

        &
        &  %
        &  %
        &  %
        &  %
        &  %
        (0.03) &  %
        (0.76)  &  %
        (-1.28) &  %
        (-0.69) &  %
        (-2.10) \\ %

        \multirow{2}{*}{Entangled} &
        \multirow{2}{*}{100}   &  %
        \multirow{2}{*}{81.25}   &  %
        \multirow{2}{*}{15.54}   &  %
        \multirow{2}{*}{60.94}   &  %
        \multirow{2}{*}{62.76}   &  %
        98.84                  &  %
        95.26                 &  %
        93.10                  &  %
        56.81                  &   %
        73.45                  \\ %

        &
        &  %
        &  %
        &  %
        &  %
        &  %
        (0.58) &  %
        (1.22)  &  %
        (3.08) &  %
        (3.84) &  %
        (6.33) \\ %

      \bottomrule
    \end{tabularx}
  \end{threeparttable}
\end{table}

To build \wmmodel, we watermarked the DNN models trained on the five datasets by
leveraging each algorithm, yielding a total of \nummodels target DNN models (5
datasets $\times$ \numwm schemes). Note that each \wmmodel should maintain its
classification accuracy and emit the predefined target labels for given key
images.

Table~\ref{tab:repro} shows the recall rate of watermark key images and accuracy
for the test instances on \wmmodel.
The second to the sixth columns summarize the trigger set recall of
\wmmodel across datasets, the fraction of the watermark
key images that are correctly classified as their target labels.
Most \wmmodel correctly remember their trigger sets and classify key samples
with a high recall of over 99\%.
The seventh to the eleventh columns describe the test accuracy for \wmmodel as
well as the magnitude of drops in test accuracy in comparison to the
corresponding models without any watermark. We observe that most \wmmodel
preserve their test accuracy after watermarking, showing only a slight drop of
within 4\%.

However, the models watermarked using \wmpassport and \wmentangled yielded relatively low
trigger set recall levels %
compared to the other target models. Note that we used the original \wmpassport
and \wmentangled implementation, resulting in no differences between our
implementation and that of the authors. Since \wmpassport utilizes abstract
images as its key images, we believe that this result stems from using a
different set of abstract key images. For the \wmentangled models, our results
accord with the trigger set recall levels reported by
Jia~\etal~\cite{jia:usec:2021}. We believe that the initial trigger set recalls
of the \wmentangled models are already too low for ownership claims.
In the remaining sections, we evaluate the presented attacks using these
watermarked models.

\subsection{Obscuring \owner's Ownership}

\label{ss:eval-obscuring}

An adversary seeking to obscure \owner's ownership attempts to thwart \owner's
ownership verification process.
For this, the adversary can employ fine-tuning, model stealing, evasion, or
parameter pruning attacks against \wmmodel, thus generating \advmodel with a low
\owner's trigger set recall.
At the same time, the test accuracy of \advmodel should not drop
significantly as the adversary needs to host a functional service by
leveraging \advmodel.
We now evaluate each attack in this category assuming both non-adaptive and
adaptive adversaries.

\subsubsection{Fine-tuning Attack}
\label{ss:eval-finetuning}

\newcolumntype{L}[1]{>{\raggedright\arraybackslash}m{#1}}
\newcolumntype{Y}[1]{>{\centering\arraybackslash}m{#1}}

\begin{table}
  \begin{threeparttable}
    \scriptsize
    \centering
    \renewcommand{\arraystretch}{1.15}
    \caption{Trigger set recall (\%) of \advmodel after fine-tuning attacks.}
    \label{tab:fine-tuning}
    \setlength\tabcolsep{0.125cm}
    \begin{tabularx}{\columnwidth}{
        L{1.10cm}|
        Y{0.50cm}
        Y{0.50cm}
        Y{0.50cm}
        Y{0.50cm}
        Y{0.50cm}|
        Y{0.50cm}
        Y{0.50cm}
        Y{0.50cm}
        Y{0.50cm}
        Y{0.50cm}
      }
      \toprule

        &
        \multicolumn{5}{c}{\textbf{Non-adaptive Attack}} &
        \multicolumn{5}{c}{\textbf{Adaptive Attack}} \\

      \cmidrule(lr){2-6}
      \cmidrule(lr){7-11}

        &
        \textbf{MN} &
        \textbf{GT} &
        \textbf{C10} &
        \textbf{TI} &
        \textbf{C100} &
        \textbf{MN} &
        \textbf{GT} &
        \textbf{C10} &
        \textbf{TI} &
        \textbf{C100} \\

      \midrule
        Content &
        \undersixty
        57.02   &  %
        \underten
        0.36   &  %
        \underforty
        24.54   &  %
        \underten
        0.00    &  %
        13.20   &  %
        \underforty
        39.74   &  %
        \undersixty
        53.96   &  %
        \underforty
        26.92   &  %
        \underten
        0.00    &  %
        75.20   \\ %

        Noise &
        \underten
        5.93   &  %
        84.46   &  %
        99.14   &  %
        \underten
        0.40    &  %
        93.80   &  %
        \underten
        0.36   &  %
        \underten
        5.63   &  %
        \underten
        3.78   &  %
        \underten
        0.40    &  %
        10.20   \\ %

        Unrelated &
        99.34   &  %
        100   &  %
        99.10   &  %
        99.40   &  %
        92.80   &  %
        \underforty
        32.76   &  %
        99.77   &  %
        \undertwenty
        17.26   &  %
        \undertwenty
        15.80   &  %
        0.00   \\ %

        Mark &
        \undersixty
        40.28   &  %
        \underten
        8.95   &  %
        \underten
        3.86   &  %
        \underten
        5.64    &  %
        \underten
        2.29   &  %
        \undertwenty
        19.77   &  %
        \underforty
        25.87   &  %
        \underten
        8.02   &  %
        \underten
        1.25    &  %
        1.46   \\ %

        Abstract &
        \undersixty
        51.00   &  %
        \undersixty
        51.00   &  %
        \undereighty
        60.00   &  %
        100     &  %
        26.00   &  %
        \undersixty
        45.00   &  %
        83.00   &  %
        \undersixty
        54.00   &  %
        100     &  %
        23.00   \\ %

        Adv &
        \underforty
        35.00   &  %
        \undereighty
        79.00   &  %
        \underforty
        24.00   &  %
        \undereighty
        66.00   &  %
        13.00   &  %
        \undertwenty
        14.00   &  %
        \underten
        8.00   &  %
        \undertwenty
        12.00  &  %
        \underten
        6.00    &  %
        \underten
        2.00   \\ %

        Passport &
        \undertwenty
        14.00   &  %
        \undersixty
        43.00   &  %
        14.00   &  %
        74.00   &  %
        3.00   &  %
        \undertwenty
        13.00   &  %
        \undersixty
        43.00   &  %
        17.00   &  %
        75.00   &  %
        3.00   \\ %

        Encoder &
        \undertwenty
        20.00   &  %
        \underten
        4.08   &  %
        \undertwenty
        20.00   &  %
        \underten
        7.00    &  %
        8.00   &  %
        \undertwenty
        17.00   &  %
        \underten
        7.14   &  %
        \underforty
        20.60   &  %
        \underten
        1.60    &  %
        5.60   \\ %

        Exp &
        \underten
        6.00   &  %
        \underten
        0.00   &  %
        \underten
        1.00   &  %
        5.00    &  %
        1.00   &  %
        \underten
        7.00   &  %
        \underten
        0.00   &  %
        \underten
        2.00   &  %
        9.00    &  %
        0.00   \\ %

        DeepSigns &
        \undertwenty
        11.00   &  %
        \underten
        1.00   &  %
        \underten
        8.00   &  %
        \underten
        1.00    &  %
        \underten
        0.00   &  %
        \undertwenty
        11.00   &  %
        \underten
        1.00   &  %
        \undertwenty
        12.00   &  %
        \underten
        0.00    &  %
        \underten
        2.00   \\ %

        Entangled &
        99.81 &  %
        \underforty
        27.34 &  %
        \underten
        4.57 &  %
        \underten
        2.68 &  %
        \undersixty
        40.89 &  %
        97.42 &  %
        \underforty
        33.59 &  %
        \underten
        1.48 &  %
        \underten
        4.69  &  %
        23.18 \\ %

      \bottomrule
    \end{tabularx}
  \end{threeparttable}
\end{table}

\noindent\textbf{\\Non-adaptive attack.}
A non-adaptive adversary tunes \wmmodel on a dataset that does not include any
key images, thus constructing another model \advmodel.
As \adv has access to 50\% of a test set, we leveraged this set to fine-tune
\wmmodel; however, using this set alone might decrease the test accuracy of
\advmodel. Therefore, we also used an extra set of images when simulating
fine-tuning attacks.
Similar to the method proposed by Chen~\etal~\cite{chen:asiaccs:2021}, we
collected arbitrary images and labeled each of them with the output of \wmmodel.
For \wmmodel trained on MNIST, we collected all images from the Fashion-MNIST
dataset~\cite{xiao:arxiv:2017}.
We took images from CIFAR-100 for fine-tuning the GTSRB, CIFAR-10, and
TinyImageNet models.
For \wmmodel trained on CIFAR-100, we collected images from CIFAR-10.

\noindent\textbf{Adaptive attack.}
In addition to these training instances, the adaptive adversary harnesses the
autoencoder-generated key images to make \wmmodel unlearn \owner's trigger set.
Recall from \S\ref{ss:adaptive} that this adversary collects $x'$, which is
designed to resemble \owner's key images that trigger $y_t$.
Therefore, the adversary assigns a random label other than $y_t$ to $x'$ and
provides this pair as a training instance for fine-tuning attacks,
expecting that \advmodel will interpret \owner's key images as the
adversary-chosen random classes.
For training each autoencoder, it takes 3--12 minutes for each class, depending
on the dataset.

When fine-tuning \wmmodel, we optimized \wmmodel using
Adam~\cite{kingma:iclr:2015} and trained \wmmodel for 10 epochs. We fixed the
learning rates at 0.01, 0.0001, and 0.0005 for MNIST, TinyImageNet, and the
other datasets, respectively, except for one case: for \wmmodel trained on the
CIFAR datasets and watermarked using \wmpassport, we used a fixed learning rate
of 0.0001. We selected these learning rates after exploratory experiments.

Table~\ref{tab:fine-tuning} presents the trigger set recall of \advmodel, which
is the resulting model after conducting fine-tuning attacks on \wmmodel.
Note that it is challenging to set a minimum trigger set recall sufficient to
prove \owner's ownership. Thus, in the table, we colored the cells of vulnerable
watermarking schemes that rendered a trigger set recall lower than the
threshold varying from 10\% to 80\%. The gradations represent the extent to
which the model is vulnerable to the attacks. We excluded \advmodel that
exhibited over a 5\% drop in test accuracy because these models do not suffice
for the adversary's goal of hosting functional services.
In Supplemental Material~5~\cite{lee:tdsc-sup:2022}, we include an expanded version of
Table~\ref{tab:fine-tuning} that displays the test accuracies of \advmodel as
well as their trigger set recalls.

The left half of the table shows the results for the non-adaptive attacks. When
we set 10\% as the minimum trigger set recall to prove ownership, the
non-adaptive fine-tuning attacks only worked against the 19 target models.
However, the number of vulnerable target models jumped to 36 in total when the
minimum requirement was set to 80\%.
Interestingly, all models watermarked using \wmmark and DeepSigns were
vulnerable to this fine-tuning attack. On the other hand, all models watermarked
with \wmunrelated were robust, demonstrating trigger set recalls of over 92\%
for all the datasets.

The right half of the table summarizes the results for the adaptive attacks.
Note in the table that the adaptive attacks further destroyed schemes that were
robust to the non-adaptive attacks.
For instance, \wmnoise and \wmunrelated models exhibited significant trigger
set recall drops.
On the other hand, the GTSRB model with \wmunrelated was robust to the adaptive
attack. Recall that we collected an extra set of images when conducting
fine-tuning attacks to preserve the test accuracy. We observed that this extra
set hindered unlearning the trigger set of the GTSRB model with \wmunrelated.
Specifically, we found that if we exclude this set when launching the attacks,
we can successfully decrease the trigger set recall down to 0\% without loss of
test accuracy.
Considering that the adversary can either include or exclude this extra set when
launching fine-tuning attacks, we conclude that \wmunrelated is also vulnerable
to the adaptive fine-tuning attack.
\begin{figure}[t]
\centering
\includegraphics[width=0.98\columnwidth]{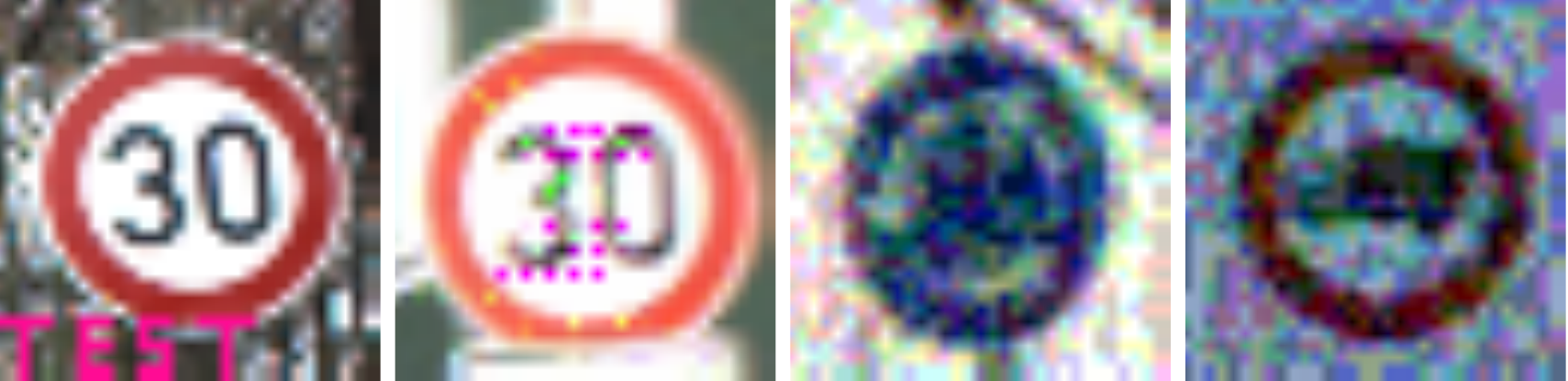}
\caption{Pairs of \owner's key image and an autoencoder-generated key image. The
left and the right halves show examples of \wmcontent and \wmnoise,
respectively.}
\vspace{-1.5em}
\label{fig:reconst-sample}
\end{figure}

We also observed that adaptive attacks are worse than non-adaptive attacks in
several cases. For instance, the non-adaptive and adaptive fine-tuning attacks
against the GTSRB model with \wmcontent reduce the trigger set recall to 0.36\%
and 53.96\%, respectively.
We carefully analyzed these cases and founded that the performance of
autoencoders used for conducting adaptive attacks varies considerably based on
the target watermarking algorithms.
Figure~\ref{fig:reconst-sample} shows examples of \owner's key images and
autoencoder-generated images.
Note from the figure that the autoencoders trained to simulate \wmnoise is
capable of generating an image that has a distribution similar to that of
\owner's key images. On the other hand, the autoencoders reported poor
performance against a watermarking scheme embedding contents, which requires
more sophisticated trigger generation.
We thus conclude that this performance difference has affected the success of
adaptive attacks in removing the embedded trigger sets.

We emphasize that none of the previous studies conducted the adaptive attack
even though they are mostly vulnerable to this attack.
Furthermore, although eight out of the \numwm watermarking algorithms had already
been evaluated against fine-tuning attacks in previous
studies~\cite{zhang:asiaccs:2018, adi:usec:2018, fan:nips:2019, li:acsac:2019, jia:usec:2021,
rouhani:asplos:2019}, our analysis reveals that many of them are still
vulnerable.
This implies that fine-tuning attacks that previous studies have conducted were
too weak to construct a meaningful upper bound of their watermarking algorithms.
Therefore, we recommend that follow-up studies evaluate their schemes against
fine-tuning attacks with sufficiently strong settings and demonstrate the extent
to which their watermarks can withstand attacks without being removed.
We further investigate various attack settings that can affect the strength of
fine-tuning attacks in Supplemental Material~3~\cite{lee:tdsc-sup:2022}.

\subsubsection{Model Stealing Attack}
\label{ss:eval-stealing}

In model stealing attacks, an adversary does not have enough training
instances to train a new model from scratch. Thus, the adversary prepares a set
of arbitrary images and leverages \wmmodel to label these images. The
adversary then trains \advmodel from scratch on these instances, thereby copying
\wmmodel's functionality except for the capability of remembering the trigger set.
We consider both non-adaptive and adaptive adversaries in evaluating the target
models against model stealing attacks.

\noindent\textbf{Non-adaptive attack.}
To collect training instances, we took the same approach as we did for
fine-tuning attacks (\S\ref{ss:eval-finetuning}).
For training, we selected \advmodel to have the same model structure as
\wmmodel. Note that the adversary knows the exact structure of \wmmodel
because \wmmodel is already in her hands. We performed model stealing attacks by
training this new model from scratch with the collected dataset.

\noindent\textbf{Adaptive attack.}
The adaptive adversary in this attack scenario also leverages the trigger set
created with the autoencoders to preclude \advmodel from learning \owner's
trigger set.
The adversary prepares training instances in the exact same way as the
adaptive adversary in fine-tuning attacks and appends them to the
training set for training \advmodel.
Because the adversary feeds $x'$ with a random label to \advmodel for its
training, this new model cannot learn \owner's trigger set.

\newcolumntype{L}[1]{>{\raggedright\arraybackslash}m{#1}}
\newcolumntype{Y}[1]{>{\centering\arraybackslash}m{#1}}

\begin{table}
  \begin{threeparttable}
    \scriptsize
    \centering
    \renewcommand{\arraystretch}{1.15}
    \caption{Trigger set recall (\%) of \advmodel after model stealing attacks.}
    \label{tab:stealing}
    \setlength\tabcolsep{0.125cm}
    \begin{tabularx}{\columnwidth}{
        L{1.10cm}|
        Y{0.50cm}
        Y{0.50cm}
        Y{0.50cm}
        Y{0.50cm}
        Y{0.50cm}|
        Y{0.50cm}
        Y{0.50cm}
        Y{0.50cm}
        Y{0.50cm}
        Y{0.50cm}
      }
      \toprule

        &
        \multicolumn{5}{c}{\textbf{Non-adaptive Attack}} &
        \multicolumn{5}{c}{\textbf{Adaptive Attack}} \\

      \cmidrule(lr){2-6}
      \cmidrule(lr){7-11}

        &
        \textbf{MN} &
        \textbf{GT} &
        \textbf{C10} &
        \textbf{TI} &
        \textbf{C100} &
        \textbf{MN} &
        \textbf{GT} &
        \textbf{C10} &
        \textbf{TI} &
        \textbf{C100} \\

      \midrule

        Content &
        82.94  &  %
        \underten
        0.00   &  %
        \underten
        2.04   &  %
        1.00   &  %
        0.80   &  %
        \underforty
        28.37  &  %
        \underten
        0.05  &  %
        \underten
        1.18  &  %
        1.80    & %
        0.40  \\  %

        Noise &
        \underten
        0.21   &  %
        \undersixty
        59.19  &  %
        \underten
        3.60   &  %
        2.20   &  %
        45.60  &  %
        \underten
        0.09  &  %
        \underten
        6.76  &  %
        \underten
        4.04  &  %
        0.20    & %
        87.60  \\  %

        Unrelated &
        99.76  &  %
        100    &  %
        95.26  &  %
        54.60  &  %
        0.00   &  %
        \underforty
        34.94  &  %
        100  &  %
        \underten
        9.02  &  %
        20.20   & %
        0.00  \\  %

        Mark &
        \undertwenty
        11.57  &  %
        \underten
        5.10   &  %
        \underten
        2.32   &  %
        0.66   &  %
        0.90   &  %
        \underten
        7.66  &  %
        \underten
        7.27  &  %
        \underten
        6.81  &  %
        0.81    & %
        1.46  \\  %

        Abstract &
        \undersixty
        41.00  &  %
        \underforty
        35.00  &  %
        \underforty
        24.00  &  %
        65.00  &  %
        2.00   &  %
        \underforty
        39.00  &  %
        \undersixty
        48.00  &  %
        \underforty
        27.00  &  %
        66.00   & %
        2.00  \\  %

        Adv &
        \underforty
        23.00  &  %
        \undereighty
        70.00  &  %
        \underten
        8.00   &  %
        31.00  &  %
        11.00  &  %
        \undertwenty
        17.00  &  %
        \underten
        0.00  &  %
        \undertwenty
        16.00 &  %
        17.00   & %
        1.00  \\  %

        Passport &
        \underten
        7.00   &  %
        \underforty
        34.00  &  %
        19.00  &  %
        60.00  &  %
        2.00   &  %
        \underten
        7.00  &  %
        \underforty
        37.00  &  %
        16.00  &  %
        57.00   & %
        1.00  \\  %

        Encoder &
        \underten
        9.67   &  %
        \underten
        2.55   &  %
        \undertwenty
        12.60  &  %
        0.80   &  %
        1.80   &  %
        \underten
        9.33  &  %
        \underten
        2.30  &  %
        \undertwenty
        13.20  &  %
        1.00    & %
        1.40  \\  %

        Exp &
        \underten
        1.00   &  %
        \underten
        0.00   &  %
        \underten
        2.00   &  %
        0.00   &  %
        0.00   &  %
        \underten
        4.00  &  %
        \underten
        0.00  &  %
        \underten
        2.00  &  %
        0.00    & %
        2.00  \\  %

        DeepSigns &
        \underten
        10.00  &  %
        \underten
        3.00   &  %
        \underten
        6.00   &  %
        0.00   &  %
        1.00   &  %
        \underten
        6.00  &  %
        \underten
        2.00  &  %
        \undertwenty
        11.00  &  %
        0.00    & %
        0.00  \\  %

        Entangled &
        99.93  &  %
        \undereighty
        75.78  &  %
        52.02  &  %
        8.04   &  %
        35.68  &  %
        96.20  &  %
        \underforty
        25.00  &  %
        37.62  &  %
        6.70   & %
        21.35  \\  %

      \bottomrule
    \end{tabularx}
  \end{threeparttable}
\end{table}

Table~\ref{tab:stealing} summarizes the experimental results of model stealing
attacks. We shaded (in red) the cells according to the same criteria as we did for
Table~\ref{tab:fine-tuning}.
The left half of the table presents the results from the non-adaptive
attacks. Overall, the target models underwent a more drastic trigger set recall
drop compared to fine-tuning attacks. However, the target models trained on the
CIFAR-100 and TinyImageNet dataset experienced a huge test accuracy drop along
with a trigger set recall drop, which indicates that our model stealing attacks
are ineffective against target models with a large number of classes.
As a result, assuming the minimal trigger set recall to be 10\%, 16 target
models were vulnerable to this attack. When we consider 80\% as the minimal
requirement, 26 models failed to verify \owner's ownership.
Among the \numwm watermarking schemes, \wmmodel with \wmexp and DeepSigns were
the most vulnerable models, showing a trigger set recall of below 10\% after the
attacks.

The right half of the table shows the results of the adaptive model stealing
attacks. Considering 10\% as the required minimal trigger set recall, the
watermarks embedded in 18 out of the \nummodels target models were destroyed by
the adaptive attack.
We also note that the trigger set recalls further decreased in most target
models compared to the non-adaptive attack.
Moreover, when we raise the bar to 80\%, all \numwm watermarking schemes were
broken by this attack.
Although \wmunrelated was robust against the non-adaptive model stealing attack,
it was destroyed by the adaptive attack.

Note that most target models were vulnerable to the non-adaptive and adaptive
model stealing attacks. This means that the current watermark evaluation practice
does not consider real-world threats properly.
We stress that researchers should evaluate robustness against the complete set
of attacks, including model stealing attacks, and raise the bar of watermarking
schemes' robustness with aggressive evaluation that considers an adaptive
adversary.

\subsubsection{Parameter Pruning Attack}
\label{ss:eval-pruning}

\newcolumntype{L}[1]{>{\raggedright\arraybackslash}m{#1}}
\newcolumntype{Y}[1]{>{\centering\arraybackslash}m{#1}}

\begin{table}
  \begin{threeparttable}
    \scriptsize
    \centering
    \renewcommand{\arraystretch}{1.15}
    \caption{Trigger set recall (\%) of \advmodel after parameter pruning
    attacks.}
    \label{tab:pruning}
    \setlength\tabcolsep{0.125cm}
    \begin{tabularx}{\columnwidth}{
        L{1.10cm}|
        Y{0.50cm}
        Y{0.50cm}
        Y{0.50cm}
        Y{0.50cm}
        Y{0.50cm}|
        Y{0.50cm}
        Y{0.50cm}
        Y{0.50cm}
        Y{0.50cm}
        Y{0.50cm}
      }
      \toprule
        &
        \multicolumn{5}{c}{\textbf{Non-adaptive Attack}} &
        \multicolumn{5}{c}{\textbf{Adaptive Attack}} \\

      \cmidrule(lr){2-6}
      \cmidrule(lr){7-11}

        &
        \textbf{MN} &
        \textbf{GT} &
        \textbf{C10} &
        \textbf{TI} &
        \textbf{C100} &
        \textbf{MN} &
        \textbf{GT} &
        \textbf{C10} &
        \textbf{TI} &
        \textbf{C100} \\

      \midrule
        Content &
        99.87   &  %
        100     &  %
        100     &  %
        99.80   &  %
        100     &  %
        \undereighty
        64.95   &  %
        100   &  %
        100   &  %
        0.00  &  %
        100   \\  %

        Noise &
        100     &  %
        100     &  %
        100     &  %
        99.00   &  %
        100     &  %
        97.29   &  %
        \underten
        0.27   &  %
        100   &  %
        0.00  &  %
        58.20   \\  %

        Unrelated &
        99.38   &  %
        100     &  %
        100     &  %
        70.20   &  %
        100     &  %
        \undertwenty
        17.22   &  %
        0.81   &  %
        90.98   &  %
        100     &  %
        4.00   \\  %

        Mark &
        97.40   &  %
        99.64   &  %
        99.64   &  %
        41.65   &  %
        94.63   &  %
        \undereighty
        69.03   &  %
        97.03   &  %
        96.97   &  %
        51.80   &  %
        69.96   \\  %

        Abstract &
        \undereighty
        73.00   &  %
        100     &  %
        100     &  %
        74.00   &  %
        100     &  %
        \undereighty
        78.00   &  %
        95.00   &  %
        97.00   &  %
        3.00    &  %
        98.00   \\  %

        Adv &
        91.00   &  %
        100     &  %
        100     &  %
        41.00   &  %
        100     &  %
        96.00   &  %
        7.00   &  %
        97.00  &  %
        12.00  &  %
        94.00   \\  %

        Passport &
        \undereighty
        80.00   &  %
        100     &  %
        \undereighty
        71.00   &  %
        99.00   &  %
        87.00   &  %
        84.00   &  %
        94.00   &  %
        82.00   &  %
        94.00   &  %
        91.00   \\  %

        Encoder &
        96.50   &  %
        96.94   &  %
        99.20   &  %
        80.80   &  %
        98.60   &  %
        99.00   &  %
        91.07   &  %
        98.20   &  %
        0.30    &  %
        92.80   \\  %

        Exp &
        92.00   &  %
        100     &  %
        100     &  %
        4.00    &  %
        100     &  %
        92.00   &  %
        99.00   &  %
        98.00   &  %
        0.00    &  %
        97.00   \\  %

        DeepSigns &
        \underforty
        39.00   &  %
        89.00   &  %
        100     &  %
        2.00    &  %
        98.00   &  %
        81.00   &  %
        98.00   &  %
        99.00   &  %
        12.00   &  %
        78.00   \\  %

        Entangled &
        100     &  %
        83.59   &  %
        \undertwenty
        17.81   &  %
        \undereighty
        70.09   &  %
        \undereighty
        61.98   &  %
        99.95   &  %
        \undereighty
        71.88   &  %
        \underten
        5.33    &  %
        2.68    &  %
        57.81   \\  %

      \bottomrule
    \end{tabularx}
  \end{threeparttable}
\end{table}

The non-adaptive and adaptive adversaries in parameter pruning attacks attempt
to prune the parameters of \wmmodel.

\noindent\textbf{Non-adaptive attack.}
To erase \owner's watermark, the non-adaptive adversary prunes $p$\% of the
smallest parameters in \wmmodel, thus building a new model \advmodel.

\noindent\textbf{Adaptive attack.}
In adaptive pruning attacks, the adversary identifies parameters that contribute
to the classification of \owner's trigger set by leveraging the
autoencoder-generated trigger set and then removes those parameters.
Specifically, the adversary observes the differences between the neuron
activations of \wmmodel when $x$ and $x'$ are given. Note that the neurons that
render different behaviors between these images can be regarded as trigger
set-related.
The adversary thus prunes $p$\% of parameters that showed the greatest
differences. After pruning, \advmodel becomes non-reactive to \owner's trigger
set.
When pruning parameters, we only considered parameters that belong to the fully
connected layers.

Table~\ref{tab:pruning} presents the trigger set recall of \advmodel after
parameter pruning attacks.
We evaluated the effect of this attack with six different values of $p$: 5, 10, 20,
40, 60, and 80. Among the results for the six different $p$ values, we only show
the results that reported the lowest trigger set recall with a test accuracy
drop of less than 5\%.
We colored the cells according to the same criteria that we set for
Table~\ref{tab:fine-tuning}.
The left half shows the trigger set recall after the non-adaptive attacks.  In
general, we found that the watermarking schemes are robust against this attack,
which accords with the experimental results of previous
studies~\cite{zhang:asiaccs:2018, merrer:nca:2019, fan:nips:2019,
namba:asiaccs:2019, rouhani:asplos:2019, jia:usec:2021}.
Seven out of \nummodels target models showed a trigger set recall of less than 80\%;
only two models were weak against this attack when we considered 40\% as the minimal
recall required to prove ownership.

The right half of the table summarizes the experimental results after the
adaptive pruning attacks. The adaptive pruning attacks were not as strong as
other adaptive attacks; however, the adaptive attack damaged five target models
that were robust to the non-adaptive attacks.
Furthermore, note that the \wmunrelated models tend to demonstrate a significant
drop in the trigger set recall, although they experience non-trivial test
accuracy drops as well (see Supplemental Material~5~\cite{lee:tdsc-sup:2022}).
These results suggest the necessity of our adaptive pruning attacks
against the existing watermarking algorithms.

\subsubsection{Evasion Attack}
\label{ss:eval-evasion}

The goal of \adv in performing an evasion attack is to distinguish queries
that have key images from normal queries. Once a key image is identified,
the adversary may return random labels to drop the trigger set recall, thus
obscuring \owner's ownership.

To assess \adv's capability of distinguishing key images from regular images,
we trained autoencoders for each class of images with 50\% of a test set. For
instance, we prepared a total of 100 autoencoders for CIFAR-100.
We then evaluated whether the trained autoencoders could output an image similar to the input image.
Note that these autoencoders are able to reconstruct normal images well but fail with key
images as the autoencoders are trained on regular images.
To decide whether the autoencoders fail to reconstruct given images, we
computed three metrics, i.e., $L_1$ norms, $L_2$ norms, and Jensen-Shannon
divergence, between the input and output images as in the approach of
\cite{meng:ccs:2017}.

Specifically, given an image, we query \wmmodel and record the output class. We
then reconstruct the image with the autoencoder of the output class and compute
the metrics. If all three metrics computed from the image are lower than
the thresholds, we consider the given image to be a normal one. We set the
thresholds such that false-positive rates are at most 0.1\% on the set of
images used for training the autoencoders.

\newcolumntype{L}[1]{>{\raggedright\arraybackslash}m{#1}}
\newcolumntype{Y}[1]{>{\centering\arraybackslash}m{#1}}

\begin{table}
  \begin{threeparttable}
    \scriptsize
    \centering
    \renewcommand{\arraystretch}{1.2}
    \caption{Trigger set detection accuracies when performing evasion attacks
    against the target models. A table cell in the red background represents a
    vulnerable model that enables \adv to detect the trigger set with an
    accuracy of over 85\%, and the gradations represent the extent to which the
    model is vulnerable to evasion attacks.}
    \label{tab:evasion}
    \begin{tabularx}{\columnwidth}{
        L{1.5cm}|
        Y{0.95cm}
        Y{0.95cm}
        Y{0.95cm}
        Y{0.95cm}
        Y{0.95cm}
      }
      \toprule
        &
        \multicolumn{5}{c}{\textbf{Detection Acc. (\%)}} \\

      \cmidrule(lr){2-6}

        &
        \textbf{MNIST} &
        \textbf{GTSRB} &
        \textbf{C10} &
        \textbf{TI} &
        \textbf{C100} \\

      \midrule

        Content &
        \overninetyfive
        98.31   &  %
        \overninetyfive
        97.57   &  %
        \overninetyfive
        98.62   &  %
        79.60   &  %
        \overeightyfive
        89.20   \\ %

        Noise &
        \overninetyfive
        98.38   &  %
        \overninetyfive
        97.75   &  %
        \overninetyfive
        99.64   &  %
        \overninety
        90.20   &  %
        \overninety
        90.30   \\ %

        Unrelated &
        \overninetyfive
        98.32   &  %
        \overninetyfive
        97.91   &  %
        49.73   &  %
        39.40   &  %
        \overeightyfive
        86.70   \\ %

        Mark &
        \overninetyfive
        98.29   &  %
        \overninety
        93.18   &  %
        \overninety
        93.88   &  %
        75.20   &  %
        \overeightyfive
        86.38   \\ %

        Abstract &
        \overninetyfive
        99.50   &  %
        \overeightyfive
        87.00   &  %
        67.50   &  %
        60.00   &  %
        81.00   \\ %

        Adv &
        \overninetyfive
        99.50   &  %
        \overninetyfive
        97.50   &  %
        \overninetyfive
        100     &  %
        \overninety
        93.50   &  %
        \overninety
        91.00   \\ %

        Passport &
        \overninetyfive
        99.50   &  %
        \overeightyfive
        89.00   &  %
        60.50   &  %
        50.00   &  %
        78.50   \\ %

        Encoder &
        \overninety
        94.75   &  %
        \overeightyfive
        85.59   &  %
        \overeightyfive
        89.40   &  %
        79.35   &  %
        \overeightyfive
        87.10   \\ %

        Exp &
        62.50   &  %
        \overeightyfive
        85.50   &  %
        78.00   &  %
        55.50   &  %
        82.00   \\ %

        DeepSigns &
        \overninetyfive
        100     &  %
        \overninetyfive
        97.00   &  %
        \overninetyfive
        100     &  %
        \overninety
        93.50   &  %
        \overninety
        93.00   \\ %

        Entangled &
        \overninetyfive
        98.55   &  %
        \overninety
        92.19   &  %
        52.47   &  %
        84.49   &  %
        \overeightyfive
        87.76   \\ %

      \bottomrule
    \end{tabularx}
  \end{threeparttable}
\end{table}

Table~\ref{tab:evasion} summarizes the detection accuracies of evasion attacks.
We balanced the number of key images and regular images when measuring the
detection accuracy so that the baseline detection accuracy is 50\%. These regular images were taken from the training set so that they would not
overlap with the images used to train the autoencoders. A high detection accuracy
implies that the adversary can successfully reduce the trigger set recall
without losing test accuracy.

As shown in the table, detection accuracies against the TinyImageNet models
are lower than those against the other models. Since we train an autoencoder for
each class, the number of training instances to train each autoencoder becomes
extremely limited (e.g., 25 images) when attacking the TinyImageNet models.
Nevertheless, note in the table that 38 target models out of \nummodels can successfully
evade the verification process as they reported at least 85\% detection
accuracies.
This is not surprising as only three out of the \numwm previous studies have
considered evasion attacks in their adversarial evaluation.
Among the three previous studies that considered evasion attacks, \wmexp is robust
against this attack, as shown in the table.
This is because it takes key images from exactly the same distribution as the
regular images used for training \wmexp.
However, \wmencoder was vulnerable to evasion attacks in our settings, even
though a previous study~\cite{li:acsac:2019} demonstrated its robustness against this attack
scenario.
That is, the previous study took a naive approach to conduct evasion attacks so
that it failed to demonstrate a meaningful upper bound on its robustness against
evasion attacks (see Supplemental Material~3~\cite{lee:tdsc-sup:2022}).

\subsection{Claiming Ownership by \adv}
\label{ss:eval-claiming}

The goal of a non-adaptive adversary claiming her ownership is to cause a
stalemate in the ownership dispute game against \owner. To simulate this
adversary, all prior research has considered a scenario where an adversary
conducts single ownership piracy or ambiguity attacks.
However, there exists an obvious solution to identify the authentic owner; thus
the adversary necessarily loses in this game (\S\ref{ss:shortcomings}).

With this in mind, we propose a new attack scenario that incorporates watermark removal
attacks within ownership claiming attacks. Specifically, we consider a novel
scenario where the adversary first removes \owner's watermark and then implants
\adv's watermark, thus claiming the ownership of a new model that only holds \adv's
watermark.
Among the watermark removal attacks, we chose models constructed via
model stealing attacks as a base for ownership piracy and ambiguity attacks due
to model stealing attacks' outstanding performance in removing watermarks
(recall \S\ref{ss:eval-stealing}).

Recall that the non-adaptive adversary in these attack scenarios claims
ownership based on her own trigger set. In other words, the adversary needs to
choose one watermarking algorithm to prepare her trigger set. For this, we
assumed that \adv prepares her trigger set using \wmunrelated. Hence,
\wmunrelated becomes the basis of \adv's fraudulent ownership claim of the
resulting model \advmodel.

\subsubsection{Ownership Piracy Attack}
\label{ss:eval-piracy}

To perform piracy attacks, the adversary follows the same procedures and
settings as fine-tuning attacks. The only difference is that \adv also appends
her trigger set to the dataset of a fine-tuning attacker. With this dataset,
\adv fine-tunes a watermark-removed model to embed her trigger set.

\newcolumntype{L}[1]{>{\raggedright\arraybackslash}m{#1}}
\newcolumntype{Y}[1]{>{\centering\arraybackslash}m{#1}}

\begin{table}
  \begin{threeparttable}
    \scriptsize
    \centering
    \caption{Trigger set recalls of \advmodel after ownership piracy attacks.
    Numbers in parentheses denote the differences of trigger set recalls between
    \adv and \owner.}
    \label{tab:piracy}
    \setlength\tabcolsep{0.07cm}
    \begin{tabularx}{\columnwidth}{
        L{1.10cm}|
        Y{0.72cm}
        Y{0.72cm}
        Y{0.82cm}
        Y{0.82cm}
        Y{0.72cm}|
        Y{0.48cm}
        Y{0.48cm}
        Y{0.48cm}
        Y{0.48cm}
        Y{0.48cm}
      }
      \toprule
        &
        \multicolumn{5}{c}{\textbf{\adv's Recall (\%)}} &
        \multicolumn{5}{c}{\textbf{\owner's Recall (\%)}} \\

      \cmidrule(lr){2-6}
      \cmidrule(lr){7-11}

        &
        \textbf{MN} &
        \textbf{GT} &
        \textbf{C10} &
        \textbf{TI} &
        \textbf{C100} &
        \textbf{MN} &
        \textbf{GT} &
        \textbf{C10} &
        \textbf{TI} &
        \textbf{C100} \\

      \midrule

        \multirow{2}{*}{Content} &
        \overtwenty
        86.00   &  %
        \overeighty
        99.00   &  %
        100   &  %
        100   &  %
        100   &  %
        \multirow{2}{*}{49.76}   &  %
        \multirow{2}{*}{0.05}   &  %
        \multirow{2}{*}{0.28}   &  %
        \multirow{2}{*}{1.00}   &  %
        \multirow{2}{*}{0.20}   \\ %

        &
        \overtwenty
        (36.24) &  %
        \overeighty
        (98.96) &  %
        (99.72) &  %
        (99.00) &  %
        (99.80) & %
        &  %
        &  %
        &  %
        &  %
        \\

        \multirow{2}{*}{Noise} &
        \overeighty
        92.00   &  %
        \oversixty
        99.00   &  %
        100   &  %
        95.83 &  %
        100   &  %
        \multirow{2}{*}{0.09}   &  %
        \multirow{2}{*}{24.05}   &  %
        \multirow{2}{*}{2.80}   &  %
        \multirow{2}{*}{0.20}   &  %
        \multirow{2}{*}{22.00}   \\ %

        &
        \overeighty
        (91.91) &  %
        \oversixty
        (74.95) &  %
        (97.20) &  %
        (95.63) &  %
        (78.00) & %
        &  %
        &  %
        &  %
        &  %
        \\

        \multirow{2}{*}{Unrelated} &
        \overeighty
        96.00   &  %
        \overtwenty
        98.00   &  %
        94.00   &  %
        100.00  &  %
        99.00   &  %
        \multirow{2}{*}{9.28}   &  %
        \multirow{2}{*}{60.72}   &  %
        \multirow{2}{*}{50.52}   &  %
        \multirow{2}{*}{22.40}   &  %
        \multirow{2}{*}{0.00}   \\ %

        &
        \overeighty
        (86.72) &  %
        \overtwenty
        (37.28) &  %
        (43.48) &  %
        (77.60) &  %
        (99.00) & %
        &  %
        &  %
        &  %
        &  %
        \\

        \multirow{2}{*}{Mark} &
        \oversixty
        87.00   &  %
        \overeighty
        99.00   &  %
        \overeighty
        98.00   &  %
        22.46 &  %
        100   &  %
        \multirow{2}{*}{11.54}   &  %
        \multirow{2}{*}{5.62}   &  %
        \multirow{2}{*}{3.06}   &  %
        \multirow{2}{*}{1.03}   &  %
        \multirow{2}{*}{0.68}   \\ %

        &
        \oversixty
        (75.46) &  %
        \overeighty
        (93.38) &  %
        \overeighty
        (94.94) &  %
        (21.44) &  %
        (99.32) & %
        &  %
        &  %
        &  %
        &  %
        \\

        \multirow{2}{*}{Abstract} &
        \oversixty
        89.00   &  %
        \oversixty
        99.00   &  %
        98.00   &  %
        100   &  %
        100   &  %
        \multirow{2}{*}{28.00}   &  %
        \multirow{2}{*}{32.00}   &  %
        \multirow{2}{*}{17.00}   &  %
        \multirow{2}{*}{61.00}   &  %
        \multirow{2}{*}{0.00}   \\ %

        &
        \oversixty
        (61.00) &  %
        \oversixty
        (67.00) &  %
        (81.00) &  %
        (39.00) &  %
        (100)   & %
        &  %
        &  %
        &  %
        &  %
        \\

        \multirow{2}{*}{Adv} &
        \overforty
        75.00   &  %
        \oversixty
        91.00   &  %
        \overeighty
        98.00   &  %
        100   &  %
        100   &  %
        \multirow{2}{*}{34.00}   &  %
        \multirow{2}{*}{12.00}   &  %
        \multirow{2}{*}{6.00}   &  %
        \multirow{2}{*}{30.00}   &  %
        \multirow{2}{*}{7.00}   \\ %

        &
        \overforty
        (41.00) &  %
        \oversixty
        (79.00) &  %
        \overeighty
        (92.00) &  %
        (70.00) &  %
        (93.00) & %
        &  %
        &  %
        &  %
        &  %
        \\

        \multirow{2}{*}{Passport} &
        \overeighty
        94.00   &  %
        0.00   &  %
        0.00   &  %
        4.00  &  %
        15.00   &  %
        \multirow{2}{*}{5.00}   &  %
        \multirow{2}{*}{6.00}   &  %
        \multirow{2}{*}{10.00}   &  %
        \multirow{2}{*}{48.00}   &  %
        \multirow{2}{*}{1.00}   \\ %

        &
        \overeighty
        (89.00) &  %
        (-6.00) &  %
        (-10.00) &  %
        (-44.00) &  %
        (14.00) & %
        &  %
        &  %
        &  %
        &  %
        \\

        \multirow{2}{*}{Encoder} &
        \overeighty
        98.00   &  %
        \overeighty
        100   &  %
        100   &  %
        100   &  %
        100   &  %
        \multirow{2}{*}{10.00}   &  %
        \multirow{2}{*}{2.30}   &  %
        \multirow{2}{*}{11.80}   &  %
        \multirow{2}{*}{0.50}   &  %
        \multirow{2}{*}{1.00}   \\ %

        &
        \overeighty
        (88.00) &  %
        \overeighty
        (97.70) &  %
        (88.20) &  %
        (99.50) &  %
        (99.00) & %
        &  %
        &  %
        &  %
        &  %
        \\

        \multirow{2}{*}{Exp} &
        \oversixty
        70.00   &  %
        \overeighty
        98.00   &  %
        98.00   &  %
        100   &  %
        100   &  %
        \multirow{2}{*}{0.00}   &  %
        \multirow{2}{*}{0.00}   &  %
        \multirow{2}{*}{2.00}   &  %
        \multirow{2}{*}{0.00}   &  %
        \multirow{2}{*}{0.00}   \\ %

        &
        \oversixty
        (70.00) &  %
        \overeighty
        (98.00) &  %
        (96.00) &  %
        (100) &  %
        (100)   & %
        &  %
        &  %
        &  %
        &  %
        \\

        \multirow{2}{*}{DeepSigns} &
        \overeighty
        93.00   &  %
        \overeighty
        100   &  %
        \overeighty
        99.00   &  %
        100   &  %
        100   &  %
        \multirow{2}{*}{5.00}   &  %
        \multirow{2}{*}{2.00}   &  %
        \multirow{2}{*}{7.00}   &  %
        \multirow{2}{*}{1.00}   &  %
        \multirow{2}{*}{0.00}   \\ %

        &
        \overeighty
        (88.00) &  %
        \overeighty
        (98.00) &  %
        \overeighty
        (92.00) &  %
        (99.00) &  %
        (100)   & %
        &  %
        &  %
        &  %
        &  %
        \\

        \multirow{2}{*}{Entangled} &
        100   &  %
        \oversixty
        100   &  %
        \overeighty
        99.93 &  %
        95.31 &  %
        98.05 &  %
        \multirow{2}{*}{100}   &  %
        \multirow{2}{*}{32.03}   &  %
        \multirow{2}{*}{12.08}   &  %
        \multirow{2}{*}{1.34}   &  %
        \multirow{2}{*}{0.00}   \\ %

        &
        (0.00)   &  %
        \oversixty
        (67.97)   &  %
        \overeighty
        (87.86)   &  %
        (93.97)   &  %
        (98.05)   & %
        &  %
        &  %
        &  %
        &  %
        \\

      \bottomrule
    \end{tabularx}
  \end{threeparttable}
\end{table}

Table~\ref{tab:piracy} presents the trigger set recalls of \adv and \owner after
the attack.
We compare these two trigger set recalls because \adv in this attack insists
that \advmodel only contains \adv's trigger set and has never been trained on
\owner's trigger set. That is, \adv aims to demonstrate that \adv's trigger set
recall is high, whereas \owner's trigger set recall is low.
We thus show the differences between the trigger set recall rates of \adv and \owner in
parentheses. We shaded (in red) the target models that showed a difference greater than a
threshold varying from 20\% to 80\%. The gradations illustrate the magnitude of
each trigger recall difference. As we did for all other attacks, we excluded
target models with a test accuracy drop of over 5\%.

Since we removed \owner's watermark before embedding \adv's watermark, \adv's
trigger set acquired a dominant position over that of \owner from 24 target
models.
Furthermore, from 21 target models, \adv's trigger set recall surpassed \owner's
by at least 60\%.
These results imply that \adv can successfully take the ownership of those target
models, claiming that those models only contain \adv's trigger set.
Considering these results, we suggest future researchers prove their algorithms'
robustness against piracy attacks based on our new scenario.

\subsubsection{Ambiguity Attack}
\label{ss:eval-ambiguity}

\newcolumntype{L}[1]{>{\raggedright\arraybackslash}m{#1}}
\newcolumntype{Y}[1]{>{\centering\arraybackslash}m{#1}}

\begin{table}
  \begin{threeparttable}
    \scriptsize
    \centering
    \caption{Trigger set recalls of \wmmodel after ambiguity attacks. Numbers
    in parentheses denote the differences of trigger set recalls between \adv
    and \owner.}
    \label{tab:ambiguity}
    \setlength\tabcolsep{0.07cm}
    \begin{tabularx}{\columnwidth}{
        L{1.10cm}|
        Y{0.72cm}
        Y{0.72cm}
        Y{0.82cm}
        Y{0.82cm}
        Y{0.72cm}|
        Y{0.48cm}
        Y{0.48cm}
        Y{0.48cm}
        Y{0.48cm}
        Y{0.48cm}
      }
      \toprule
        &
        \multicolumn{5}{c}{\textbf{\adv's Recall (\%)}} &
        \multicolumn{5}{c}{\textbf{\owner's Recall (\%)}} \\

      \cmidrule(lr){2-6}
      \cmidrule(lr){7-11}

        &
        \textbf{MN} &
        \textbf{GT} &
        \textbf{C10} &
        \textbf{TI} &
        \textbf{C100} &
        \textbf{MN} &
        \textbf{GT} &
        \textbf{C10} &
        \textbf{TI} &
        \textbf{C100} \\

      \midrule

        \multirow{2}{*}{Content} &
        100   &  %
        \overeighty
        98.00 &  %
        \overeighty
        100   &  %
        100   &  %
        100   &  %
        \multirow{2}{*}{82.94}   &  %
        \multirow{2}{*}{0.00}    &  %
        \multirow{2}{*}{2.04}    &  %
        \multirow{2}{*}{1.00}   &  %
        \multirow{2}{*}{0.80}    \\  %

        &
        (17.06) &  %
        \overeighty
        (98.00) &  %
        \overeighty
        (97.96) &  %
        (99.00) &  %
        (99.20) & %
        &  %
        &  %
        &  %
        &  %
        \\

        \multirow{2}{*}{Noise} &
        \overeighty
        100   &  %
        \overtwenty
        98.00 &  %
        \overeighty
        100   &  %
        100   &  %
        100   &  %
        \multirow{2}{*}{0.21}    &  %
        \multirow{2}{*}{59.19}   &  %
        \multirow{2}{*}{3.60}    &  %
        \multirow{2}{*}{2.20}   &  %
        \multirow{2}{*}{45.60}   \\  %

        &
        \overeighty
        (99.79) &  %
        \overtwenty
        (38.81)  &  %
        \overeighty
        (96.40) &  %
        (97.80) &  %
        (54.40) & %
        &  %
        &  %
        &  %
        &  %
        \\

        \multirow{2}{*}{Unrelated} &
        98.00   &  %
        100   &  %
        100   &  %
        99.00 &  %
        97.00   &  %
        \multirow{2}{*}{99.76}  &  %
        \multirow{2}{*}{100}   &  %
        \multirow{2}{*}{95.26}   &  %
        \multirow{2}{*}{54.60}   &  %
        \multirow{2}{*}{0.00}    \\  %

        &
        (-1.76) &  %
        (0.00)  &  %
        (4.74)  &  %
        (44.40) &  %
        (97.00)  & %
        &  %
        &  %
        &  %
        &  %
        \\

        \multirow{2}{*}{Mark} &
        \overeighty
        100    &  %
        \overeighty
        100    &  %
        \overeighty
        100    &  %
        100   &  %
        94.00    &  %
        \multirow{2}{*}{11.57}  &  %
        \multirow{2}{*}{5.10}   &  %
        \multirow{2}{*}{2.32}   &  %
        \multirow{2}{*}{0.66}   &  %
        \multirow{2}{*}{0.90}   \\  %

        &
        \overeighty
        (88.43) &  %
        \overeighty
        (94.90) &  %
        \overeighty
        (97.68) &  %
        (99.34) &  %
        (93.10) & %
        &  %
        &  %
        &  %
        &  %
        \\

        \multirow{2}{*}{Abstract} &
        \overforty
        100   &  %
        \oversixty
        100   &  %
        \oversixty
        100   &  %
        100   &  %
        100   &  %
        \multirow{2}{*}{41.00}  &  %
        \multirow{2}{*}{35.00}  &  %
        \multirow{2}{*}{24.00}  &  %
        \multirow{2}{*}{65.00}   &  %
        \multirow{2}{*}{2.00}  \\  %

        &
        \overforty
        (59.00) &  %
        \oversixty
        (65.00) &  %
        \oversixty
        (76.00) &  %
        (35.00) &  %
        (98.00) & %
        &  %
        &  %
        &  %
        &  %
        \\

        \multirow{2}{*}{Adv} &
        \oversixty
        100   &  %
        \overtwenty
        100   &  %
        \overeighty
        100   &  %
        99.00 &  %
        100   &  %
        \multirow{2}{*}{23.00}  &  %
        \multirow{2}{*}{70.00}  &  %
        \multirow{2}{*}{8.00}  &  %
        \multirow{2}{*}{31.00}   &  %
        \multirow{2}{*}{11.00}  \\  %

        &
        \oversixty
        (77.00) &  %
        \overtwenty
        (30.00) &  %
        \overeighty
        (92.00) &  %
        (68.00) &  %
        (89.00)  & %
        &  %
        &  %
        &  %
        &  %
        \\

        \multirow{2}{*}{Passport} &
        \overeighty
        100   &  %
        48.00 &  %
        0.00  &  %
        0.00  &  %
        41.00 &  %
        \multirow{2}{*}{7.00} &  %
        \multirow{2}{*}{34.00} &  %
        \multirow{2}{*}{19.00} &  %
        \multirow{2}{*}{60.00}   &  %
        \multirow{2}{*}{2.00} \\  %

        &
        \overeighty
        (93.00)   &  %
        (14.00)  &  %
        (-19.00) &  %
        (-60.00) &  %
        (39.00) &  %
        &  %
        &  %
        &  %
        &  %
        \\

        \multirow{2}{*}{Encoder} &
        \overeighty
        100   &  %
        \overeighty
        99.00 &  %
        \overeighty
        100   &  %
        98.00 &  %
        100   &  %
        \multirow{2}{*}{9.67}  &  %
        \multirow{2}{*}{2.55}  &  %
        \multirow{2}{*}{12.60}  &  %
        \multirow{2}{*}{0.80}   &  %
        \multirow{2}{*}{1.80}  \\  %

        &
        \overeighty
        (90.33) &  %
        \overeighty
        (96.45) &  %
        \overeighty
        (87.40) &  %
        (97.20) &  %
        (98.20) & %
        &  %
        &  %
        &  %
        &  %
        \\

        \multirow{2}{*}{Exp} &
        \overforty
        48.00     &  %
        0.00     &  %
        \overeighty
        100     &  %
        97.00 &  %
        0.00     &  %
        \multirow{2}{*}{1.00}  &  %
        \multirow{2}{*}{0.00}   &  %
        \multirow{2}{*}{2.00}   &  %
        \multirow{2}{*}{0.00}   &  %
        \multirow{2}{*}{0.00}   \\  %

        &
        \overforty
        (47.00) &  %
        (0.00) &  %
        \overeighty
        (98.00) &  %
        (97.00) &  %
        (0.00) & %
        &  %
        &  %
        &  %
        &  %
        \\

        \multirow{2}{*}{DeepSigns} &
        \overeighty
        100   &  %
        \overeighty
        100   &  %
        \overeighty
        100   &  %
        100   &  %
        98.00   &  %
        \multirow{2}{*}{10.00}  &  %
        \multirow{2}{*}{3.00}   &  %
        \multirow{2}{*}{6.00}  &  %
        \multirow{2}{*}{0.00}   &  %
        \multirow{2}{*}{1.00}   \\  %

        &
        \overeighty
        (90.00) &  %
        \overeighty
        (97.00) &  %
        \overeighty
        (94.00) &  %
        (100) &  %
        (97.00) & %
        &  %
        &  %
        &  %
        &  %
        \\

        \multirow{2}{*}{Entangled} &
        100   &  %
        99.00 &  %
        97.00 &  %
        98.00 &  %
        99.00 &  %
        \multirow{2}{*}{99.93}   &  %
        \multirow{2}{*}{75.78}   &  %
        \multirow{2}{*}{52.02}   &  %
        \multirow{2}{*}{8.04}   &  %
        \multirow{2}{*}{35.68}   \\ %

        &
        (0.07)   &  %
        (23.22)   &  %
        (44.98)   &  %
        (89.96)   &  %
        (63.32)   & %
        &  %
        &  %
        &  %
        &  %
        \\

      \bottomrule
    \end{tabularx}
  \end{threeparttable}
\end{table}

Unlike the previous adversary, an adversary performing ambiguity attacks does
not implant \adv's trigger set into the watermark-removed model.  Instead, \adv
generates key images that can trigger the adversary-chosen target labels when
given to the watermark-removed model. Consequently, \adv can claim that the
resulting model only remembers \adv's trigger set.
Specifically, \adv perturbs seed images by leveraging gradient descent in order
to divert the classification of the perturbed images towards the
adversary-chosen target label~\cite{fredrikson:ccs:2015}.

\begin{figure}[t]
\centering
\includegraphics[width=0.98\columnwidth]{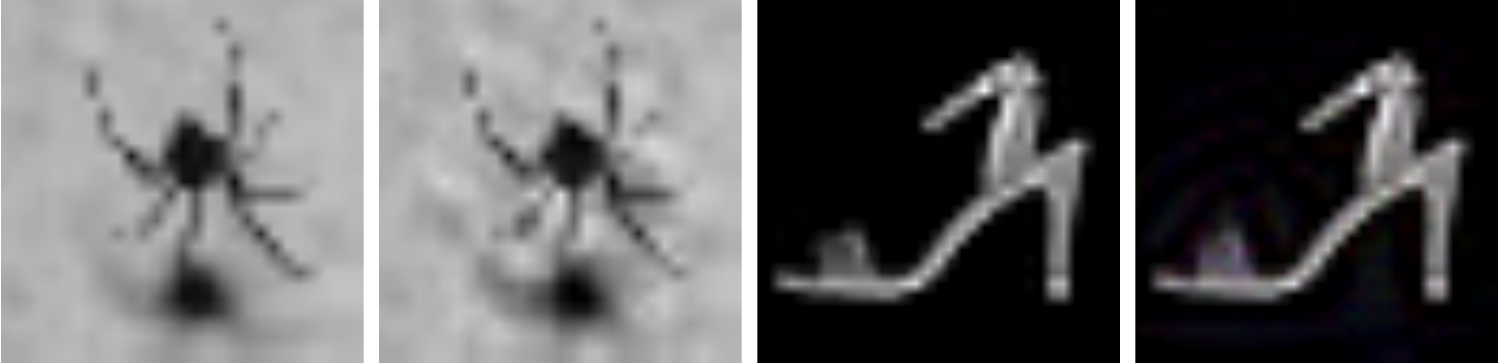}
\caption{\adv's key images (second and fourth images) generated from seed images
(first and third images) using ambiguity attacks.}
\vspace{-1.5em}
\label{fig:ambiguity-res}
\end{figure}

Table~\ref{tab:ambiguity} summarizes the ambiguity attack results. We applied
the same criteria as piracy attacks to color each cell. Since \adv
does not modify the watermark-removed model, \owner's trigger set recalls after
this attack are the same as those shown in Table~\ref{tab:stealing}.
Among \nummodels target models, \adv's trigger set recalls from 23 target models
were greater than \owner's by at least 20\%.
The second and fourth images in Figure~\ref{fig:ambiguity-res} show examples of
key images generated by ambiguity attacks targeting the \wmnoise-MNIST and
\wmcontent-GTSRB models, respectively.
We note that the $L_2$ norm of perturbations added to the
seed images is less than 0.004 on average, which implies that the added
perturbations are quasi-imperceptible.
We thus conclude that \adv can successfully claim her ownership of those models
with the created trigger set based on \wmunrelated.
As we demonstrated the effectiveness of this attack, we argue that future
research should also evaluate its algorithm against ambiguity attacks.

\section{Lessons}
\label{sec:lessons}

\newcolumntype{L}[1]{>{\raggedright\arraybackslash}m{#1}}
\newcolumntype{Y}[1]{>{\centering\arraybackslash}m{#1}}

\begin{table*}
  \begin{threeparttable}
    \footnotesize
    \centering
    \renewcommand{\arraystretch}{1.1}
    \caption{Summary of the attack results. \cmark~denotes that the attack
    succeeded against a target model watermarked with the corresponding
    algorithm, whereas \xmark~indicates that the attack failed.
    For each watermarking scheme, the successful attacks are presented in the
    order of MNIST, GTSRB, CIFAR-10, TinyImageNet, and CIFAR-100 models.}
    \label{tab:summary}
    \setlength\tabcolsep{0.105cm}
    \begin{tabularx}{\textwidth}{
        L{1.50cm}|
        Y{0.087cm}
        Y{0.087cm}
        Y{0.087cm}
        Y{0.087cm}
        Y{0.087cm}|
        Y{0.087cm}
        Y{0.087cm}
        Y{0.087cm}
        Y{0.087cm}
        Y{0.087cm}|
        Y{0.087cm}
        Y{0.087cm}
        Y{0.087cm}
        Y{0.087cm}
        Y{0.087cm}|
        Y{0.087cm}
        Y{0.087cm}
        Y{0.087cm}
        Y{0.087cm}
        Y{0.087cm}|
        Y{0.087cm}
        Y{0.087cm}
        Y{0.087cm}
        Y{0.087cm}
        Y{0.087cm}|
        Y{0.087cm}
        Y{0.087cm}
        Y{0.087cm}
        Y{0.087cm}
        Y{0.087cm}|
        Y{0.087cm}
        Y{0.087cm}
        Y{0.087cm}
        Y{0.087cm}
        Y{0.087cm}|
        Y{0.087cm}
        Y{0.087cm}
        Y{0.087cm}
        Y{0.087cm}
        Y{0.087cm}|
        Y{0.087cm}
        Y{0.087cm}
        Y{0.087cm}
        Y{0.087cm}
        Y{0.087cm}|
        Y{0.087cm}
        Y{0.087cm}
        Y{0.087cm}
        Y{0.087cm}
        Y{0.087cm}|
        Y{0.087cm}
        Y{0.087cm}
        Y{0.087cm}
        Y{0.087cm}
        Y{0.087cm}
      }
      \toprule
        \textbf{Attack (Adv.)} &
        \multicolumn{5}{c}{\rotate{\wmcontent}} &
        \multicolumn{5}{c}{\rotate{\wmnoise}} &
        \multicolumn{5}{c}{\rotate{\wmunrelated}} &
        \multicolumn{5}{c}{\rotate{\wmmark}} &
        \multicolumn{5}{c}{\rotate{\wmabstract}} &
        \multicolumn{5}{c}{\rotate{\wmadv}} &
        \multicolumn{5}{c}{\rotate{\wmpassport}} &
        \multicolumn{5}{c}{\rotate{\wmencoder}} &
        \multicolumn{5}{c}{\rotate{\wmexp}} &
        \multicolumn{5}{c}{\rotate{DeepSigns}} &
        \multicolumn{5}{c}{\rotate{\wmentangled}} \\

      \midrule
        Fine-tuning (non-adap.)&
        \cmark &  %
        \cmark &  %
        \cmark &  %
        \cmark &  %
        \xmark &  %
        \cmark &  %
        \xmark &  %
        \xmark &  %
        \cmark &  %
        \xmark &  %
        \xmark &  %
        \xmark &  %
        \xmark &  %
        \xmark &  %
        \xmark &  %
        \cmark &  %
        \cmark &  %
        \cmark &  %
        \cmark &  %
        \cmark &  %
        \cmark &  %
        \cmark &  %
        \cmark &  %
        \xmark &  %
        \xmark &  %
        \cmark &  %
        \cmark &  %
        \cmark &  %
        \cmark &  %
        \xmark &  %
        \cmark &  %
        \cmark &  %
        \xmark &  %
        \xmark &  %
        \xmark &  %
        \cmark &  %
        \cmark &  %
        \cmark &  %
        \cmark &  %
        \xmark &  %
        \cmark &  %
        \cmark &  %
        \cmark &  %
        \xmark &  %
        \xmark &  %
        \cmark &  %
        \cmark &  %
        \cmark &  %
        \cmark &  %
        \cmark &  %
        \xmark &  %
        \cmark &  %
        \cmark &  %
        \cmark &  %
        \cmark \\ %

        \rowcolor{lightgray}
        Fine-tuning (adap.)&
        \cmark &  %
        \cmark &  %
        \cmark &  %
        \cmark &  %
        \xmark &  %
        \cmark &  %
        \cmark &  %
        \cmark &  %
        \cmark &  %
        \xmark &  %
        \cmark &  %
        \xmark &  %
        \cmark &  %
        \cmark &  %
        \xmark &  %
        \cmark &  %
        \cmark &  %
        \cmark &  %
        \cmark &  %
        \xmark &  %
        \cmark &  %
        \xmark &  %
        \cmark &  %
        \xmark &  %
        \xmark &  %
        \cmark &  %
        \cmark &  %
        \cmark &  %
        \cmark &  %
        \cmark &  %
        \cmark &  %
        \cmark &  %
        \xmark &  %
        \xmark &  %
        \xmark &  %
        \cmark &  %
        \cmark &  %
        \cmark &  %
        \cmark &  %
        \xmark &  %
        \cmark &  %
        \cmark &  %
        \cmark &  %
        \xmark &  %
        \xmark &  %
        \cmark &  %
        \cmark &  %
        \cmark &  %
        \cmark &  %
        \cmark &  %
        \xmark &  %
        \cmark &  %
        \cmark &  %
        \cmark &  %
        \xmark \\ %

        Stealing (non-adap.) &
        \xmark &  %
        \cmark &  %
        \cmark &  %
        \xmark &  %
        \xmark &  %
        \cmark &  %
        \cmark &  %
        \cmark &  %
        \xmark &  %
        \xmark &  %
        \xmark &  %
        \xmark &  %
        \xmark &  %
        \xmark &  %
        \xmark &  %
        \cmark &  %
        \cmark &  %
        \cmark &  %
        \xmark &  %
        \xmark &  %
        \cmark &  %
        \cmark &  %
        \cmark &  %
        \xmark &  %
        \xmark &  %
        \cmark &  %
        \cmark &  %
        \cmark &  %
        \xmark &  %
        \xmark &  %
        \cmark &  %
        \cmark &  %
        \xmark &  %
        \xmark &  %
        \xmark &  %
        \cmark &  %
        \cmark &  %
        \cmark &  %
        \xmark &  %
        \xmark &  %
        \cmark &  %
        \cmark &  %
        \cmark &  %
        \xmark &  %
        \xmark &  %
        \cmark &  %
        \cmark &  %
        \cmark &  %
        \xmark &  %
        \xmark &  %
        \xmark &  %
        \cmark &  %
        \xmark &  %
        \xmark &  %
        \xmark \\ %

        \rowcolor{lightgray}
        Stealing (adap.)&
        \cmark &  %
        \cmark &  %
        \cmark &  %
        \xmark &  %
        \xmark &  %
        \cmark &  %
        \cmark &  %
        \cmark &  %
        \xmark &  %
        \xmark &  %
        \cmark &  %
        \xmark &  %
        \cmark &  %
        \xmark &  %
        \xmark &  %
        \cmark &  %
        \cmark &  %
        \cmark &  %
        \xmark &  %
        \xmark &  %
        \cmark &  %
        \cmark &  %
        \cmark &  %
        \xmark &  %
        \xmark &  %
        \cmark &  %
        \cmark &  %
        \cmark &  %
        \xmark &  %
        \xmark &  %
        \cmark &  %
        \cmark &  %
        \xmark &  %
        \xmark &  %
        \xmark &  %
        \cmark &  %
        \cmark &  %
        \cmark &  %
        \xmark &  %
        \xmark &  %
        \cmark &  %
        \cmark &  %
        \cmark &  %
        \xmark &  %
        \xmark &  %
        \cmark &  %
        \cmark &  %
        \cmark &  %
        \xmark &  %
        \xmark &  %
        \xmark &  %
        \cmark &  %
        \xmark &  %
        \xmark &  %
        \xmark \\ %

        Pruning (non-adap.)&
        \xmark &  %
        \xmark &  %
        \xmark &  %
        \xmark &  %
        \xmark &  %
        \xmark &  %
        \xmark &  %
        \xmark &  %
        \xmark &  %
        \xmark &  %
        \xmark &  %
        \xmark &  %
        \xmark &  %
        \xmark &  %
        \xmark &  %
        \xmark &  %
        \xmark &  %
        \xmark &  %
        \xmark &  %
        \xmark &  %
        \cmark &  %
        \xmark &  %
        \xmark &  %
        \xmark &  %
        \xmark &  %
        \xmark &  %
        \xmark &  %
        \xmark &  %
        \xmark &  %
        \xmark &  %
        \cmark &  %
        \xmark &  %
        \cmark &  %
        \xmark &  %
        \xmark &  %
        \xmark &  %
        \xmark &  %
        \xmark &  %
        \xmark &  %
        \xmark &  %
        \xmark &  %
        \xmark &  %
        \xmark &  %
        \xmark &  %
        \xmark &  %
        \cmark &  %
        \xmark &  %
        \xmark &  %
        \xmark &  %
        \xmark &  %
        \xmark &  %
        \xmark &  %
        \cmark &  %
        \cmark &  %
        \cmark \\ %

        \rowcolor{lightgray}
        Pruning (adap.) &
        \cmark &  %
        \xmark &  %
        \xmark &  %
        \xmark &  %
        \xmark &  %
        \xmark &  %
        \cmark &  %
        \xmark &  %
        \xmark &  %
        \xmark &  %
        \cmark &  %
        \xmark &  %
        \xmark &  %
        \xmark &  %
        \xmark &  %
        \cmark &  %
        \xmark &  %
        \xmark &  %
        \xmark &  %
        \xmark &  %
        \cmark &  %
        \xmark &  %
        \xmark &  %
        \xmark &  %
        \xmark &  %
        \xmark &  %
        \xmark &  %
        \xmark &  %
        \xmark &  %
        \xmark &  %
        \xmark &  %
        \xmark &  %
        \xmark &  %
        \xmark &  %
        \xmark &  %
        \xmark &  %
        \xmark &  %
        \xmark &  %
        \xmark &  %
        \xmark &  %
        \xmark &  %
        \xmark &  %
        \xmark &  %
        \xmark &  %
        \xmark &  %
        \xmark &  %
        \xmark &  %
        \xmark &  %
        \xmark &  %
        \xmark &  %
        \xmark &  %
        \cmark &  %
        \cmark &  %
        \xmark &  %
        \xmark \\ %

        Evasion &
        \cmark &  %
        \cmark &  %
        \cmark &  %
        \xmark &  %
        \cmark &  %
        \cmark &  %
        \cmark &  %
        \cmark &  %
        \cmark &  %
        \cmark &  %
        \cmark &  %
        \cmark &  %
        \xmark &  %
        \xmark &  %
        \cmark &  %
        \cmark &  %
        \cmark &  %
        \cmark &  %
        \xmark &  %
        \cmark &  %
        \cmark &  %
        \cmark &  %
        \xmark &  %
        \xmark &  %
        \xmark &  %
        \cmark &  %
        \cmark &  %
        \cmark &  %
        \cmark &  %
        \cmark &  %
        \cmark &  %
        \cmark &  %
        \xmark &  %
        \xmark &  %
        \xmark &  %
        \cmark &  %
        \cmark &  %
        \cmark &  %
        \xmark &  %
        \cmark &  %
        \xmark &  %
        \cmark &  %
        \xmark &  %
        \xmark &  %
        \xmark &  %
        \cmark &  %
        \cmark &  %
        \cmark &  %
        \cmark &  %
        \cmark &  %
        \cmark &  %
        \cmark &  %
        \xmark &  %
        \xmark &  %
        \cmark \\ %

        \rowcolor{lightgray}
        Ownership Piracy &
        \cmark &  %
        \cmark &  %
        \xmark &  %
        \xmark &  %
        \xmark &  %
        \cmark &  %
        \cmark &  %
        \xmark &  %
        \xmark &  %
        \xmark &  %
        \cmark &  %
        \cmark &  %
        \xmark &  %
        \xmark &  %
        \xmark &  %
        \cmark &  %
        \cmark &  %
        \cmark &  %
        \xmark &  %
        \xmark &  %
        \cmark &  %
        \cmark &  %
        \xmark &  %
        \xmark &  %
        \xmark &  %
        \cmark &  %
        \cmark &  %
        \cmark &  %
        \xmark &  %
        \xmark &  %
        \cmark &  %
        \xmark &  %
        \xmark &  %
        \xmark &  %
        \xmark &  %
        \cmark &  %
        \cmark &  %
        \xmark &  %
        \xmark &  %
        \xmark &  %
        \cmark &  %
        \cmark &  %
        \xmark &  %
        \xmark &  %
        \xmark &  %
        \cmark &  %
        \cmark &  %
        \cmark &  %
        \xmark &  %
        \xmark &  %
        \xmark &  %
        \xmark &  %
        \xmark &  %
        \xmark &  %
        \xmark \\ %

        Ambiguity &
        \xmark &  %
        \cmark &  %
        \cmark &  %
        \xmark &  %
        \xmark &  %
        \cmark &  %
        \cmark &  %
        \cmark &  %
        \xmark &  %
        \xmark &  %
        \xmark &  %
        \xmark &  %
        \xmark &  %
        \xmark &  %
        \xmark &  %
        \cmark &  %
        \cmark &  %
        \cmark &  %
        \xmark &  %
        \xmark &  %
        \cmark &  %
        \cmark &  %
        \cmark &  %
        \xmark &  %
        \xmark &  %
        \cmark &  %
        \cmark &  %
        \cmark &  %
        \xmark &  %
        \xmark &  %
        \cmark &  %
        \xmark &  %
        \xmark &  %
        \xmark &  %
        \xmark &  %
        \cmark &  %
        \cmark &  %
        \cmark &  %
        \xmark &  %
        \xmark &  %
        \cmark &  %
        \xmark &  %
        \cmark &  %
        \xmark &  %
        \xmark &  %
        \cmark &  %
        \cmark &  %
        \cmark &  %
        \xmark &  %
        \xmark &  %
        \xmark &  %
        \xmark &  %
        \xmark &  %
        \xmark &  %
        \xmark \\ %

      \midrule
        \textbf{\# of Succeeded Attacks} &
        6 &  %
        7 &  %
        6 &  %
        2 &  %
        1 &  %
        7 &  %
        7 &  %
        5 &  %
        3 &  %
        1 &  %
        5 &  %
        2 &  %
        2 &  %
        2 &  %
        1 &  %
        8 &  %
        7 &  %
        7 &  %
        2 &  %
        2 &  %
        9 &  %
        6 &  %
        5 &  %
        0 &  %
        0 &  %
        7 &  %
        7 &  %
        7 &  %
        3 &  %
        2 &  %
        8 &  %
        5 &  %
        1 &  %
        0 &  %
        0 &  %
        7 &  %
        7 &  %
        6 &  %
        2 &  %
        1 &  %
        6 &  %
        6 &  %
        5 &  %
        0 &  %
        0 &  %
        8 &  %
        7 &  %
        7 &  %
        3 &  %
        3 &  %
        1 &  %
        6 &  %
        4 &  %
        2 &  %
        3 \\ %

        \midrule
        \textbf{Maximum \# per Scheme} &
        \multicolumn{5}{c|}{7} &
        \multicolumn{5}{c|}{7} &
        \multicolumn{5}{c|}{5} &
        \multicolumn{5}{c|}{8} &
        \multicolumn{5}{c|}{9} &
        \multicolumn{5}{c|}{7} &
        \multicolumn{5}{c|}{8} &
        \multicolumn{5}{c|}{7} &
        \multicolumn{5}{c|}{6} &
        \multicolumn{5}{c|}{8} &
        \multicolumn{5}{c}{6} \\
      \bottomrule
    \end{tabularx}
  \end{threeparttable}
\end{table*}

We have so far conducted six different attacks along with three adaptive attacks
to evaluate the robustness of the \numwm watermarking algorithms.
Table~\ref{tab:summary} shows the number of attacks that succeeded in each
target model watermarked using the given algorithm. When tallying up the totals,
we only included results that reported a test accuracy drop of at most 5\%.
We set a trigger set recall of 80\% as the minimum threshold for claiming ownership
after conducting fine-tuning, model stealing, and parameter pruning attacks. For evasion attacks, we
considered 85\% as the minimal detection accuracy necessary to evade the verification step.

Note in the table that every watermarking algorithm is broken by at least two
presented adaptive attacks and two non-adaptive attacks.
When considering both adaptive and non-adaptive attacks, all schemes do not
demonstrate their robustness against at least five attacks.
Furthermore, six out of the 11 watermarking algorithms~\cite{zhang:asiaccs:2018, adi:usec:2018, li:acsac:2019,
fan:nips:2019, rouhani:asplos:2019, jia:usec:2021} were vulnerable to attacks against that the
authors had already evaluated.
While all the evaluated watermarking algorithms were broken, \wmunrelated
was the most robust algorithm among them.

These results highlight that all the existing trigger set-based watermarking
algorithms are not ready for real-world deployment.
We believe that the demonstrated failure to establish watermark robustness stems from
current research practice regarding how  adversarial evaluation is conducted.
We further discuss several factors that make robust watermarks (\S\ref{ss:robust-factor})
and suggestions for adversarial evaluation (\S\ref{ss:suggestions-adeval}).

\subsection{Robustness of Watermarking Algorithms}
\label{ss:robust-factor}

We analyze what makes a particular watermarking algorithm more resilient to
adversarial attacks than the others.

\noindent\textbf{Distances between key images and decision boundaries.}
Note that watermark removal attacks aim to slightly distort the decision
boundaries so that the resulting model predicts the key images as a label other
than the target label. In other words, when the key images are distant from the
decision boundaries, the target model becomes resilient to watermark removal
attacks. We thus measured the distances between key images and the decision
boundaries. Specifically, we conducted the PGD attacks~\cite{madry:iclr:2018}
against the target model and utilized the perturbation size required to modify
the prediction of key images as a distance metric. We selected two most robust
(i.e., \wmnoise and \wmunrelated) and two most vulnerable (i.e., DeepSigns and
\wmmark) watermarking schemes against fine-tuning attacks for this evaluation.
When measured over the CIFAR-10 models, we observed that the distances of
\wmnoise and \wmunrelated are greater than those of DeepSigns and \wmmark.

\noindent\textbf{Effect of target labels.}
As shown in Table~\ref{tab:summary}, \wmnoise and \wmunrelated are
more robust than the other algorithms against non-adaptive fine-tuning attacks.
Note that the key difference between these two schemes and the others is
that they allocate a single class to all key images, whereas the
remaining schemes assign different labels to each key image.
That is, the consistent labeling of these two schemes helps \wmmodel
generalize on \owner's trigger set, thus making it difficult for \adv to remove
\owner's watermark.
\swimnote{Empirical verification goes here.}

\noindent\textbf{Effect of key images.}
We observed that \wmexp is the only robust algorithm against evasion attacks.
Note that \wmexp employs images selected from the same distribution as normal
training instances for key images, while the other watermarking algorithms
use out-of-distribution images.
\sooelnote{Any empirical supports?}

Considering these factors that affect watermark robustness, we propose the
following recommendations for improving watermark robustness.
First, it is better to assign a single target label to all key images rather
than random labels.
Second, it is better to select key images from the same distribution as a
regular training set rather than from a different distribution.

\subsection{Suggestions for Adversarial Evaluation}
\label{ss:suggestions-adeval}

From our evaluations, we draw the following takeaways that future
research on designing a secure watermarking algorithm should consider. We
encourage researchers to evaluate their defenses following our suggestions
discussed herein, thus demonstrating a meaningful upper bound on their robustness.

\noindent\textbf{Apply the complete attack set.}
We found out that all the previous works were broken by already existing
attacks. They could have known this result if they have conducted a complete set
of existing state-of-the-art attacks to evaluate their algorithms.
In this regard, we suggest future research conduct at least a complete set of
state-of-the-art attacks at the time of suggesting a new approach.

Recently, several watermark removal attacks~\cite{wang:oakland:2019,
li:iclr:2021, shafieinejad:arxiv:2019, chen:asiaccs:2021} that have better
performance compared to the attacks examined herein have been recently proposed.
For instance, Chen~\etal~\cite{chen:asiaccs:2021} adopted the elastic weight
consolidation algorithm to further improve the fine-tuning attacks.
We thus recommend researchers to consider these state-of-the-art attacks when
evaluating their watermarking schemes.

\noindent\textbf{Use adaptive attacks.}
All the state-of-the-art watermarking algorithms were vulnerable to the proposed
adaptive attacks.
We believe that our adaptive attacks serve as a better baseline for demonstrating
the robustness of a target watermarking scheme.
We recommend future research consider the proposed adaptive attacks when
conducting fine-tuning, model stealing, and pruning attacks.

\noindent\textbf{Focus on attacks that obscure \owner's ownership.}
Recall from \S\ref{ss:shortcomings} that an attack scenario in which the adversary
conducts a single attack that claims her ownership is futile.
Therefore, when evaluating attacks that aim to claim \adv's ownership, one
should first launch attacks that remove \owner's watermark and then initiate the
attacks to claim \adv's ownership.

\noindent\textbf{Search for effective attack hyperparameters.}
Surprisingly, five out of the 11 evaluated watermarking algorithms were broken
by attacks that the previous studies already evaluated (recall
\S\ref{ss:eval-finetuning} and \S\ref{ss:eval-evasion}).
To avoid providing a misleading upper bound on robustness, follow-on
research must conduct strong attacks by carefully exploring hyperparameters and
adopting state-of-the-art attacks.

\noindent\textbf{Consider diverse datasets.}
Overall, the models trained on the MNIST dataset tend to be vulnerable, as shown
in Table~\ref{tab:summary}.
Interestingly, the watermarked models trained on CIFAR-100 and TinyImageNet were
robust against the presented attacks in general. This is because the conducted
attacks have always contributed to decreasing a test accuracy over 5\% (see
Supplemental Material~5~\cite{lee:tdsc-sup:2022}), which means that the presented attacks on the
CIFAR-100 and TinyImageNet models easily undermine the models' performance.
In other words, we observed that test accuracies are prone to drop
significantly after watermark removal attacks when the number of classes in a
dataset increases.
Note that it is well-known that various DNN defense algorithms showed different
levels of robustness depending on the dataset~\cite{carlini:aisec:2017}.
We thus suggest considering more datasets than the MNIST and CIFAR datasets when
evaluating watermarking algorithms.

\section{Related Work}
\label{s:related}

\noindent
\textbf{Backdoor attacks.}
There have been several studies on backdoor attacks against DNN
models~\cite{gu:arxiv:2019, chen:arxiv:2017}. In this type of attack, a user sends a
training set to the adversarial trainer to outsource the training process.
The adversary then trains a model with the received normal data as well as
images containing a backdoor trigger, e.g., a sticker with a flower. The goal of
the adversary here is to lead the model to misclassify when the
backdoor-triggering input is provided.

To mitigate backdoor attacks, researchers have proposed several
mitigation methodologies~\cite{wang:oakland:2019, chen:ijcai:2019,
liu:ccs:2019}.
DeepInspect~\cite{chen:ijcai:2019} reverse-engineers the backdoor trigger using
a conditional generative model and then fine-tunes the target model by
harnessing the generated backdoor-triggering images and their correct labels.
Wang~\etal~\cite{wang:oakland:2019} suggested another method that remedies the
target model by removing neurons that contribute to misclassifying backdoor-triggering images.

Note that these defenses are similar to our adaptive fine-tuning attacks and
adaptive pruning attacks \emph{per se}. However, their approaches are not
directly applicable to reverse-engineering key images of various trigger
set-based DNN watermarking algorithms because they only focus on
backdoor-triggering inputs created by adding a backdoor trigger to the source
images.
On the other hand, we demonstrated how an adaptive adversary generates key
images against diverse watermarking schemes.

\noindent
\textbf{Adversarial example attacks.}
DNN models are known to misclassify adversarial examples created by adding
quasi-imperceptible perturbations to normal examples~\cite{szegedy:iclr:2014,
goodfellow:iclr:2015}.
Since this finding, there has been a vast volume of research on adversarial
examples.
To mitigate this threat, Papernot~\etal~\cite{papernot:oakland:2016} proposed
defensive distillation to smooth the network gradients exploited for generating
adversarial samples.
On the other hand, MagNet~\cite{meng:ccs:2017} detects such examples at the
testing phase; it detects and reforms adversarial examples by leveraging
autoencoders trained on regular images.
However, these defenses were later broken by other strong
attacks~\cite{carlini:oakland:2017, carlini:aisec:2017, carlini:arxiv:2017}.
Adversarial training~\cite{shaham:arxiv:2016}, which improves the robustness of
DNN models by training adversarial examples with correct labels, is the current
state-of-the-art defense against adversarial example
attacks~\cite{madry:iclr:2018}.
In our study, we selected \wmadv that utilizes adversarial examples as
our target watermarking scheme and employed the approach of
MagNet~\cite{meng:ccs:2017} for evasion attacks.

\noindent
\textbf{Model stealing attacks.}
The goal of model stealing attacks, also known as model extraction attacks, is
to copy the classification performance of remote target models~\cite{tramer:usec:2016}.
Papernot~\etal~\cite{papernot:asiaccs:2017} trained a counterfeit model
as a stepping stone for creating adversarial examples of remote
target models.
Orekondy~\etal~\cite{orekondy:cvpr:2019} demonstrated that model stealing is
still possible against complex DNN models even though the
adversary does not have enough training sets and does not know the model
structure. They showed that arbitrary images downloaded from the Internet and
arbitrary models are enough to forge the target model.
PRADA~\cite{juuti:eurosp:2019} detects model stealing attempts by
analyzing incoming queries. However, this defense is inapplicable to
DNN watermarking algorithms. Note that the adversary does not have to send
remote queries because the target model is already in the hands of the adversary.
We leveraged this attack for removing \owner's watermark in the target model. In
our settings, we prepared the training set for model stealing attacks in
\S\ref{ss:eval-stealing} following the approach of \cite{orekondy:cvpr:2019}.

\section{Conclusion}
We investigate the current practice of demonstrating watermark robustness
via adversarial evaluation in the previous studies. We point out two common
flaws in their evaluations: (1) incomplete adversarial evaluation and (2)
overlooked adaptive attacks.
Taking into account these shortcomings, we evaluate the 10 trigger set-based
watermarking schemes and demonstrate that  every proposed watermarking
scheme is vulnerable to at least five presented attacks, which significantly
undermines their intended goal of proving ownership.
We conclude these failures stem from today's flawed practice in conducting
adversarial evaluation. We encourage future studies on new watermarking
algorithms to consider our guidelines presented herein to demonstrate a
meaningful upper bound of robustness against the complete set of the existing
attacks, including the proposed adaptive attacks.

\section*{Acknowledgment}
This work was supported by Institute of Information \& communications Technology
Planning \& Evaluation (IITP) grant funded by the Korea government (MSIT)
(No.2020-0-00153, Penetration Security Testing of ML Model Vulnerabilities and
Defense).

\vspace{-8em}
\begin{IEEEbiography}[{
  \includegraphics[width=1in,height=1.25in,clip,keepaspectratio]{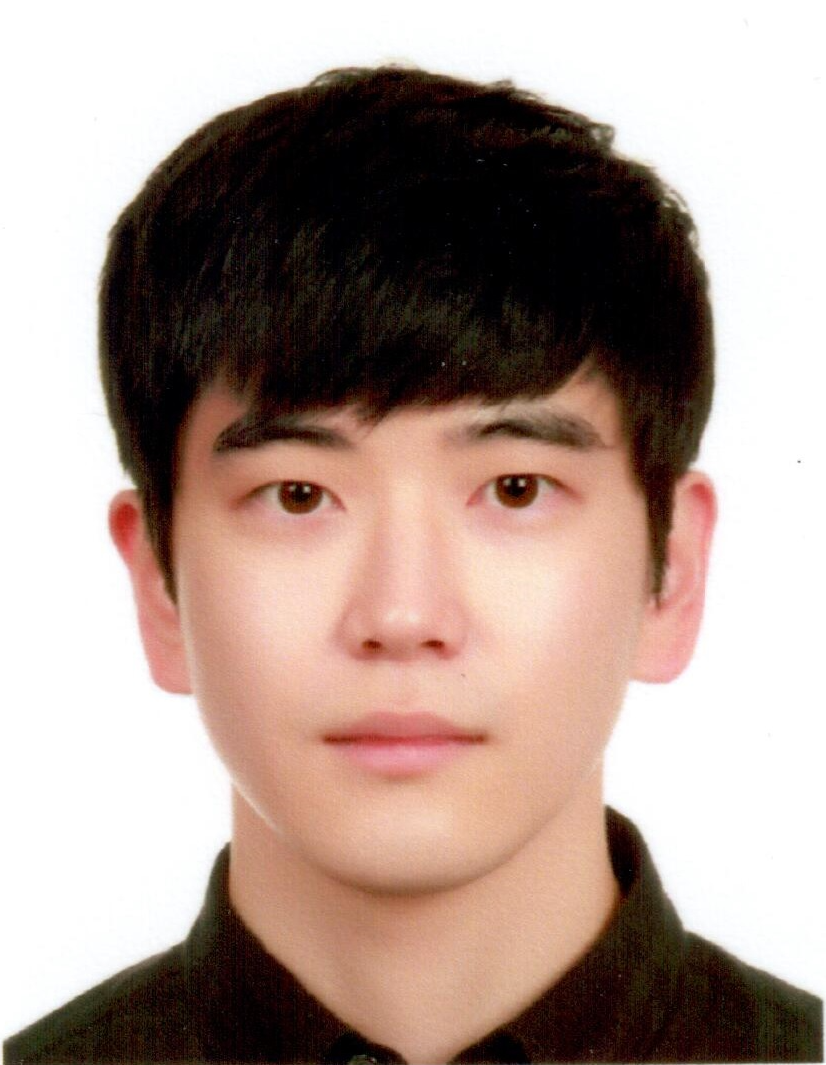}
}]{Suyoung Lee}
is a Ph.D. student at KAIST. His primary research interests lie in
the area of finding, analyzing, and patching software vulnerabilities. He has
participated in building several tools that automatically find vulnerabilities
in large software, such as web browsers and web applications.
\end{IEEEbiography}

\vspace{-8em}
\begin{IEEEbiography}[{
  \includegraphics[width=1in,height=1.25in,clip,keepaspectratio]{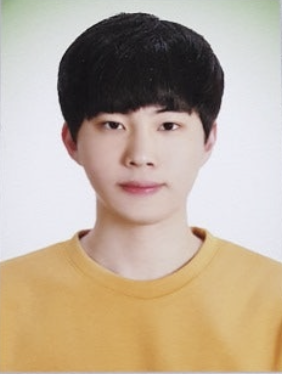}
}]{Wonho Song}
is a Master’s student at KAIST. His research mainly focuses on securing machine
learning systems.
\end{IEEEbiography}

\vspace{-8em}
\begin{IEEEbiography}[{
  \includegraphics[width=1in,height=1.25in,clip,keepaspectratio]{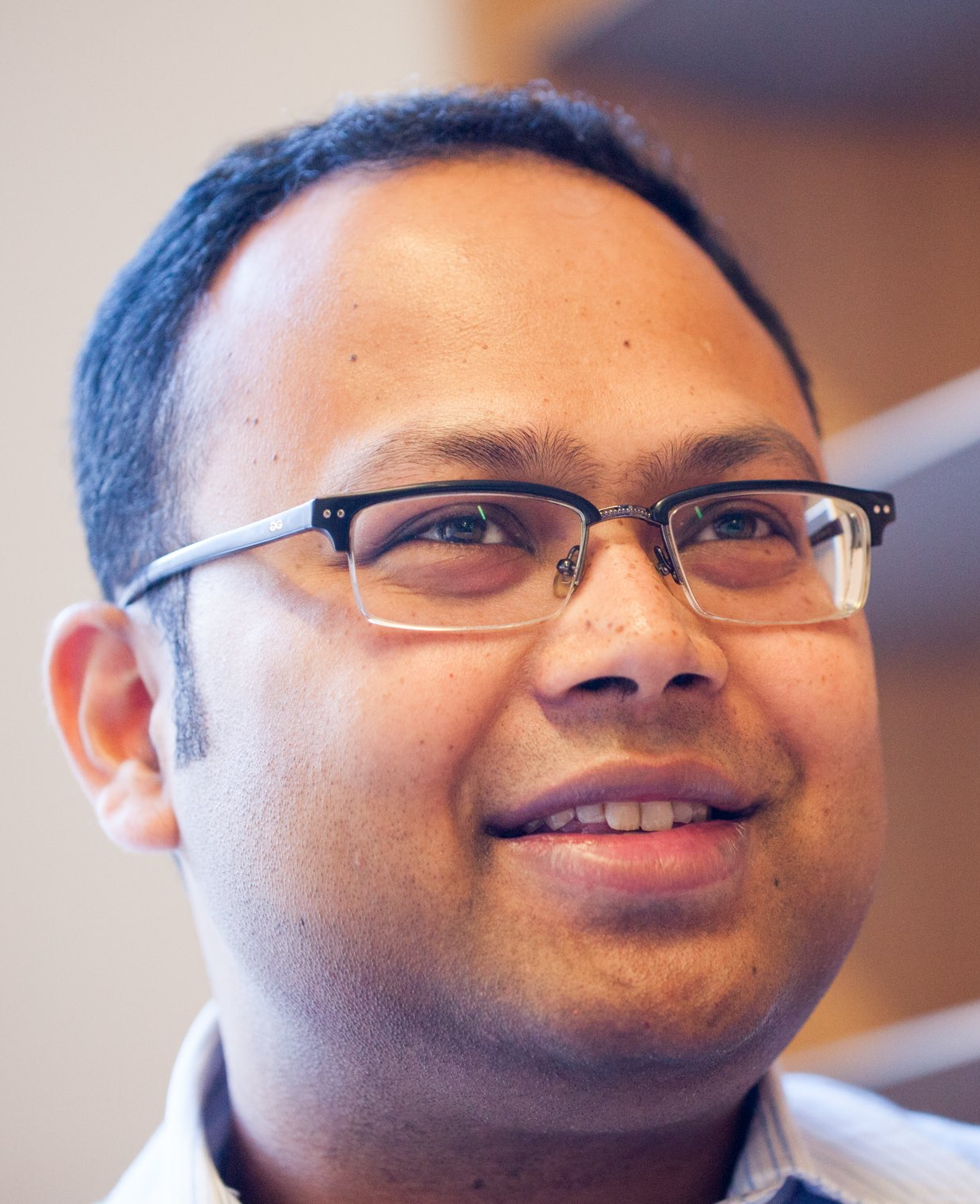}
}]{Suman Jana}
is an associate professor in the department of computer science and the data
science institute at Columbia University.  His primary research interest is at
the intersections of computer security and machine learning. His research has
received six best paper awards, a CACM research highlight, a Google faculty
fellowship, a JPMorgan Chase Faculty Research Award, an NSF CAREER award, and an
ARO young investigator award.
\end{IEEEbiography}

\vspace{-8em}
\begin{IEEEbiography}[{
  \includegraphics[width=1in,height=1.25in,clip,keepaspectratio]{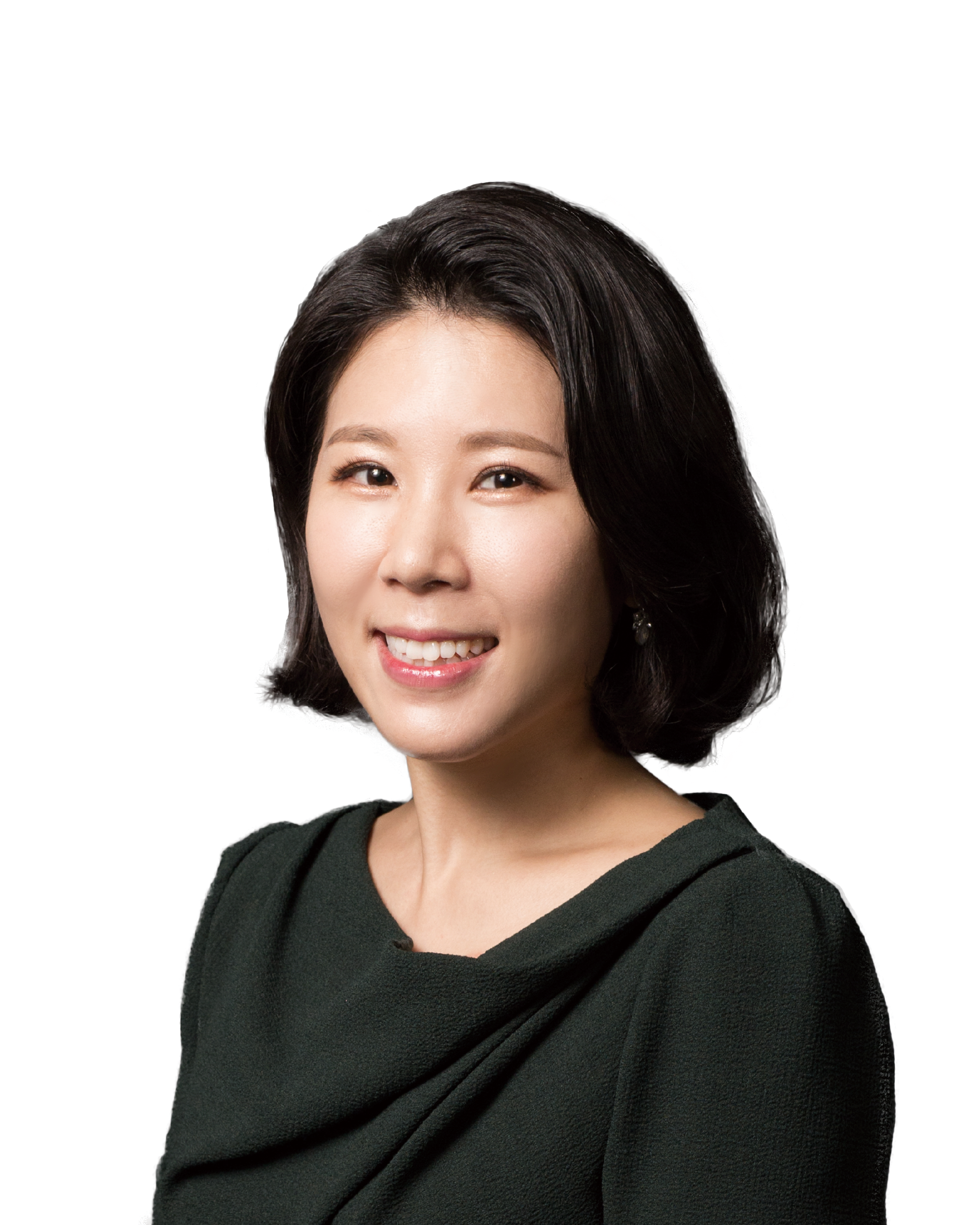}
}]{Meeyoung Cha}
is an associate professor at the Korea Advanced Institute of Science and
Technology (KAIST) in South Korea. She was a post-doctoral researcher at MPI-SWS
in Germany. Her research is on data science with an emphasis on modeling
socially relevant information propagation processes (e.g., misinformation,
poverty mapping, fraud detection, and long-tail content).  Meeyoung is a
recipient of the Korean Young Information Scientist Award, the AAAI ICWSM Test
of Time Award, and the Korean Minister of Science and ICT Award.  She worked at
Facebook’s Data Science Team as a Visiting Professor and is currently leading a
research group at the Institute for Basic Science (IBS) in Korea.
\end{IEEEbiography}

\vspace{-8em}
\begin{IEEEbiography}[{
  \includegraphics[width=1in,height=1.25in,clip,keepaspectratio]{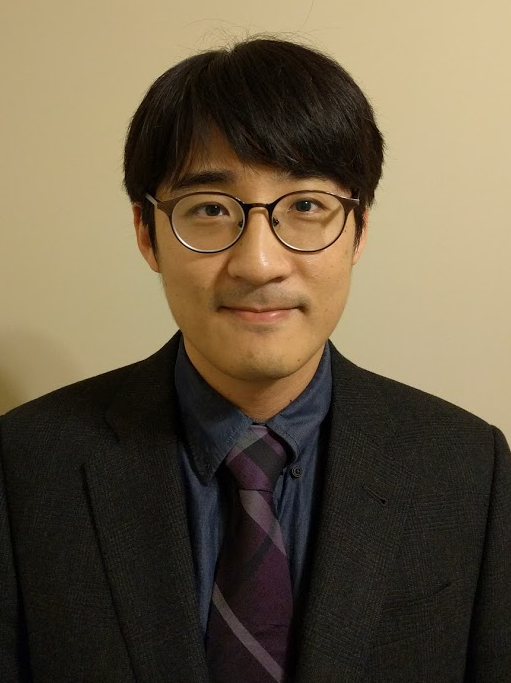}
}]{Sooel Son}
is an associate professor of School of Computing at KAIST. He received his Ph.D.
in the department of computer science at the University of Texas at Austin. He
is working on various topics regarding web security and privacy.
\end{IEEEbiography}

\end{document}